\documentclass[9pt,twocolumn,twoside]{osajnl}
\journal{optica} \setboolean{shortarticle}{false}

\usepackage{tikz}
\usetikzlibrary{arrows.meta}
\usetikzlibrary{calc}
\usepackage{amsmath}
\DeclareMathOperator*{\argmin}{arg\,min}
\usepackage{subcaption}
\usepackage{placeins}
\renewcommand{\thesubsection}{\thesection.\Alph{subsection}}

\usepackage{xcolor}

\title{Dynamic Coherent Diffractive Imaging Using \\Only a Support Constraint in the Complex Plane}

\author[1,2]{Yaocheng Tian}
\author[3]{Taichi Tsuchiya}
\author[4,5]{Yu-chen Karen Chen-Wiegart}
\author[6]{Horacio D.~Espinosa}
\author[3]{Yuichiro Kunai}
\author[1,2,7*]{George Barbastathis}

\affil[1]{Department of Mechanical Engineering, Massachusetts Institute of Technology, Cambridge, MA 02139, USA}
\affil[2]{Center for Computational Science and Engineering, Massachusetts Institute of Technology, Cambridge, MA 02139, USA}
\affil[3]{Fujikura Ltd., Tokyo 135-8512, Japan}
\affil[4]{Department of Materials Science and Chemical Engineering, Stony Brook University, Stony Brook, NY 11794, USA}
\affil[5]{National Synchrotron Light Source II, Brookhaven National Laboratory, Upton, NY 11973, USA}
\affil[6]{Department of Mechanical Engineering, Northwestern University, Evanston, IL 60208, USA}
\affil[7]{Singapore--MIT Alliance for Research and Technology (SMART) Centre, Singapore 138602, Singapore}
\affil[*]{Corresponding author: gbarb@mit.edu}

\begin{abstract}
	We show that a bounded temporal increment prior on the sample dynamics is sufficient to reconstruct a time-varying phase object from a near-field diffraction movie, under the thin-film approximation. The time evolution of the field is parameterized by a multiplicative inter-frame update factor, and a bound on its complex-plane support enforces a bounded phase increment and a passive amplitude constraint. Reconstruction of the dynamic field is thereby converted into a feasibility problem with two projection operators: a measurement-domain modulus projection and an object-domain circular-sector projection. We validate the approach experimentally using a spatial light modulator as a calibrated dynamic sample in two cases: a reaction--diffusion phase pattern with spatially expanding extent, and a growing phase pattern whose accumulated phase reaches $10\pi$. In both cases the reconstructed phase trajectory agrees well with the ground truth. We then apply the same framework, without modification, to in-situ monitoring of a photo-polymer 3D printing process, recovering the spatiotemporal phase induced by polymerization under a spatially patterned blue light. The reconstructed phase trajectory provides an observable for photo-chemical system identification and process control.
\end{abstract}

\setboolean{displaycopyright}{true}

\begin{document}
	\maketitle
	
	\section{Introduction}

Quantitative characterization of thin and often transparent samples is a recurring task in metrology and imaging. Common approaches include stylus profilometry, where a mechanical tip scans the surface line by line \cite{Whitehouse1994HandbookSurfaceMetrology}, and interferometric profilometry, where phase-shifted interferograms are processed to recover optical path length \cite{Bruning1974DigitalWavefrontInterferometer,Creath1988PhaseMeasurementInterferometry, BornWolf1999PrinciplesOfOptics}. These methods impose practical constraints. Stylus measurements require a rigid sample and a subsequent correction for finite tip geometry. Interferometric methods typically require multiple exposures and a stable environment to maintain fringe visibility and phase coherence.

Coherent diffraction imaging (CDI) removes the reference arm used in interferometric techniques and reconstructs a complex-valued field from intensity-only diffraction measurements by solving a phase retrieval problem. Without prior information, phase retrieval is non-unique. Practical CDI relies on constraints in the object domain, such as support and nonnegativity \cite{Fienup1978ReconstructionModulusFT,Fienup1982PRComparison,SeldinFienup1990Uniqueness}. Robustness can be improved by adding measurement diversity, including transverse diversity in ptychography \cite{FaulknerRodenburg2004MovableAperture,MaidenRodenburg2009PIE,GuizarSicairosFienup2008TranslationDiversity}, axial diversity via transport of intensity \cite{Teague1983TIE, WallerTianBarbastathis2010TIEHigherOrder}, and spectral diversity via chromatic aberration \cite{WallerTianBarbastathis2010ChromaticAberrationPhase}. Priors can also be learned from data; in near-field settings, learned inverses have been shown to tolerate extremely low photon counts \cite{GoyArthurLiBarbastathis2018LowPhotonPR,SinhaLeeLiBarbastathis2017LenslessDeepLearning}.

Most CDI formulations assume that the object is stationary during acquisition. Extending CDI to time-varying samples requires priors that couple frames. A representative approach uses a time-invariant region as an object-domain constraint shared across frames \cite{LoZhaoEtAl2018InSituCDI}. Other methods identify or weight such regions adaptively \cite{ShengZhang2025SerialCDIInterFrameContinuity}, or parameterize motion explicitly \cite{CaoDivekarNunezUpadhyayulaWaller2024NSTM}, but these approaches typically require either designed reference structure, an explicit motion model, or training data.

In this paper, we show that a bounded temporal increment prior on the sample dynamics is sufficient to reconstruct a time-varying phase object from a near-field diffraction movie, under the thin-film approximation with spatially coherent illumination. The field evolution is parameterized by a multiplicative inter-frame update factor, and a bound on its support in the complex plane constrains the frame-to-frame phase change and enforces a passive amplitude condition. Together with the measurement-domain modulus constraint, this yields a feasibility problem solved by projection-based iterations. We validate the method experimentally using a spatial light modulator as a calibrated dynamic sample, and demonstrate that the phase trajectory can be recovered without explicit phase unwrapping even when the accumulated phase far exceeds $2\pi$. We then apply the same framework, without modification, to \textit{in-situ} monitoring of a photo-polymer 3D printing process, recovering the spatiotemporal evolution of the induced phase as polymerization proceeds under a spatially patterned blue-color illumination.
 	\section{Inter-frame phase retrieval}\label{sec:method}

\begin{figure}[!htbp]
	\centering
	\includegraphics[width=0.45\textwidth]{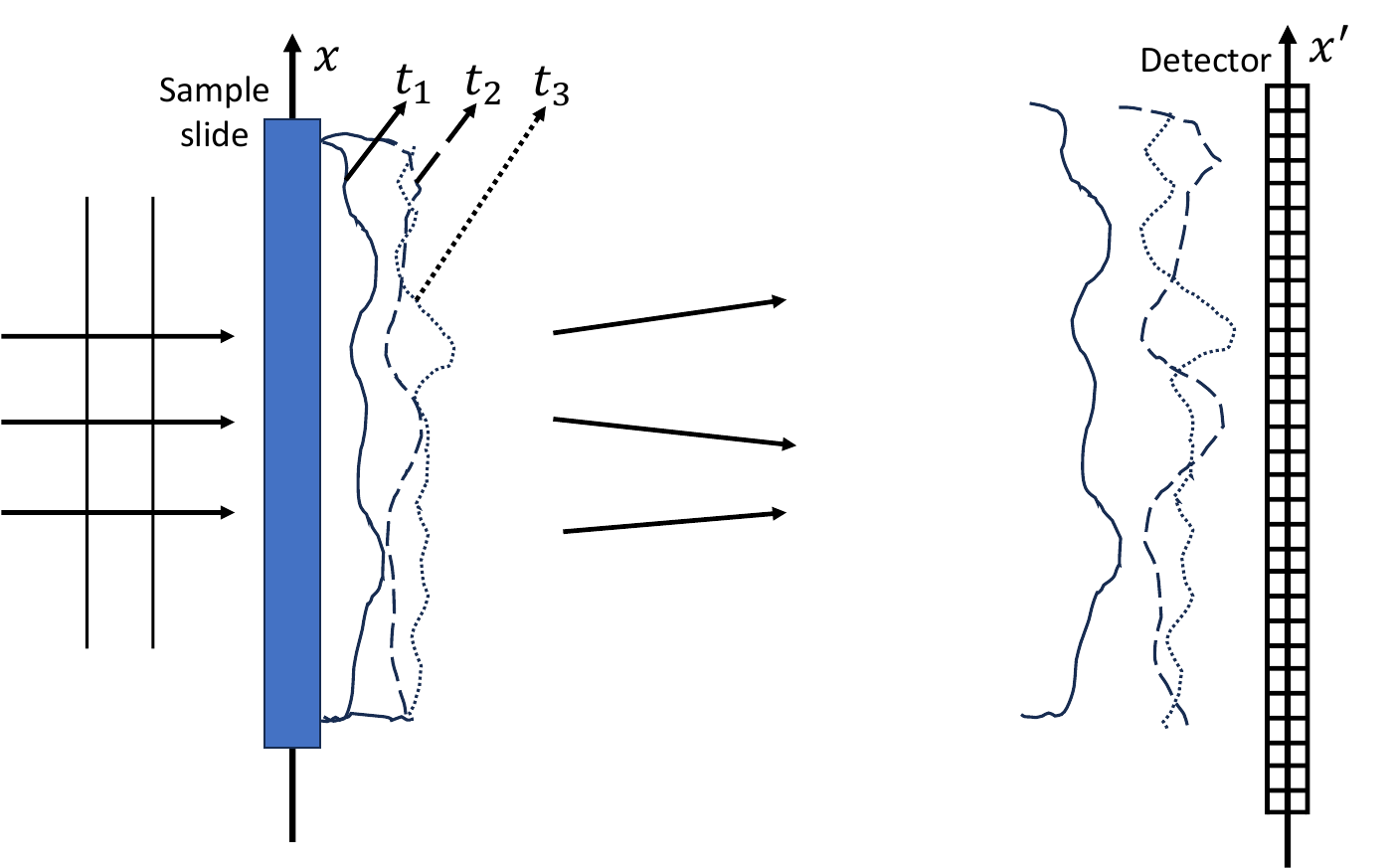}
	\caption{The lensless measurement geometry for dynamic coherent diffractive imaging.}
	\label{fig:measurement-geometry}
\end{figure}

We consider the near-field coherent diffraction measurement geometry shown in Fig.~\ref{fig:measurement-geometry}. A monochromatic, spatially coherent plane wave illuminates a dynamic sample, and a detector records the resulting Fresnel diffraction intensity $I_{\mathrm{meas}}(x',y',t)$ over time. The solid, dashed, and dotted lines on the sample plane in Fig.~\ref{fig:measurement-geometry} illustrate the wavefront of the diffracted field at three time points $t_1$, $t_2$, and $t_3$. The sample is transparent at the illumination wavelength and modifies only the phase of the illuminating field; free-space propagation then maps the spatially varying phase at the sample plane into intensity variations at the detector. The inverse problem is to recover the time-varying object-plane field $\psi(x,y,t)$, and in particular its phase $\phi(x,y,t)$, from the intensity movie $I_{\mathrm{meas}}(x',y',t)$.

We pose this reconstruction as a feasibility problem on the inter-frame field updates $\Delta\psi(x,y,t)$, which parameterize the field evolution multiplicatively. The feasibility set is the intersection of two constraint sets: a bound on the inter-frame phase increment together with a passive amplitude condition, which together confine each pixel of $\Delta\psi$ to a circular sector in the complex plane (Fig.~\ref{fig:circular-sector-in-complex-plane}); and a measurement-consistency constraint, which restricts $\Delta\psi$ to inter-frame factors whose induced diffraction intensities match the recorded movie.

Section~\ref{sec:method-multiplicative} derives the multiplicative inter-frame parameterization from the thin-film approximation. Section~\ref{sec:method-bound} derives the circular-sector constraint from a bound on the rate of change of the sample's optical path length, together with the passivity condition on the field amplitude. Section~\ref{sec:method-projectors} formulates the reconstruction as a feasibility problem and defines the two projection operators that enforce the constraints. Section~\ref{sec:algo} describes the two-stage reconstruction algorithm: first the probe field, then the inter-frame factors in a forward and backward pass through the diffraction movie. Technical derivations of the projection operators and the iteration maps are collected in the Appendix.

\subsection{Multiplicative inter-frame parameterization}\label{sec:method-multiplicative}

The Fresnel diffraction intensity recorded at the detector plane $z$ away from the sample is
\begin{equation}
	\begin{aligned}
		I_{\mathrm{meas}}(&x',y',t)
		=
		\left|
		\frac{\exp(i2\pi z/\lambda)}{i\lambda z}
		\times \right.\\
		&\left.
		\iint \psi(x,y,t)\,
		\exp\!\left[
		\frac{i\pi}{\lambda z}
		\bigl((x-x')^2+(y-y')^2\bigr)
		\right]
		\,dx\,dy
		\right|^2 .
	\end{aligned}
	\label{eq:I_meas_def}
\end{equation}
where $\lambda$ is the wavelength of the illuminating plane wave. We assume the per-frame integration time is short relative to the timescale of the sample dynamics, so motion blur within a single exposure is negligible and \eqref{eq:I_meas_def} holds at each frame time. The case of finite exposure time is beyond the scope of this paper.

Under the thin-film approximation, the field at the object plane factors as
\begin{equation}
	\psi(x,y,t) = p(x,y)\exp\!\left(i\,\phi(x,y,t)\right),
	\quad
	\phi(x,y,t) = \frac{2\pi}{\lambda}\,f(x,y,t),
	\label{eq:thin_object}
\end{equation}
where $p(x,y)$ is the illumination field, which is constant for plane-wave illumination and may more generally represent a spatially varying profile such as a focused beam in hard X-ray nanoprobe or ptychographic configurations. The projected optical path length (OPL) is
\begin{equation}
	f(x,y,t) := \int_0^{D(x,y,t)}\bigl(n(x,y,z,t)-1\bigr)\,\mathrm{d}z,
	\label{eq:OPL_def}
\end{equation}
where $D(x,y,t)$ is the sample thickness and $n(x,y,z,t)$ is the refractive index. Either or both may evolve in time: in surface-height measurement the thickness varies while the refractive index is fixed, and in photo-polymer printing between two glass slides the thickness is fixed and the refractive index evolves with the photo-chemical reaction. The framework developed below makes no assumption about which of these is the source of the dynamics; it constrains only the inter-frame phase change.

The OPL evolves between successive frames by an increment
\begin{equation}
	\Delta f(x,y,t) := f(x,y,t+\Delta t)-f(x,y,t),
	\label{eq:delta_f_def}
\end{equation}
giving the multiplicative factorization of the inter-frame field update,
\begin{align}
	\psi(x,y,t+\Delta t)
	&= p(x,y)\exp\!\left(i\frac{2\pi}{\lambda}\bigl[f(x,y,t)+\Delta f(x,y,t)\bigr]\right) \nonumber\\
	&= p(x,y)\exp\!\left(i\bigl[\phi(x,y,t)+\Delta\phi(x,y,t)\bigr]\right) \nonumber\\
	&= \psi(x,y,t)\,\Delta\psi(x,y,t),
	\label{eq:multiplicative_update_full}
\end{align}
where $\Delta\psi(x,y,t) := \exp(i\,\Delta\phi(x,y,t))$. In discrete time,
\begin{equation}
	\psi(x,y,j\Delta t) = \psi\!\left(x,y,(j-1)\Delta t\right)\,
	\Delta\psi\!\left(x,y,(j-1)\Delta t\right),
	\quad j=1,\dots,J,
	\label{eq:psi_update_timearg}
\end{equation}
and by recursion,
\begin{equation}
	\psi(x,y,j\Delta t) = \psi(x,y,0)\prod_{k=0}^{j-1}\Delta\psi(x,y,k\Delta t),
	\qquad j=1,\dots,J.
	\label{eq:psi_from_delta_product}
\end{equation}
\eqref{eq:psi_update_timearg} is Markov in $\psi(x,y,(j-1)\Delta t)$: given the current state and $\Delta\psi(x,y,(j-1)\Delta t)$, the next state is determined without reference to earlier frames.

This multiplicative form is the central parameterization of our framework. Rather than reconstructing each frame $\psi(x,y,t)$ independently and stitching them together, we reconstruct the inter-frame factors $\Delta\psi$, from which the full trajectory follows by sequential integration starting from $\psi(x,y,0)$. The next subsection shows that bounding the rate at which the sample changes between frames yields a support constraint on $\Delta\psi$ in the complex plane.

\subsection{Circular-sector constraint from bounded sample dynamics}\label{sec:method-bound}

\begin{figure}[t]
	\centering

\begin{tikzpicture}[>=stealth, scale=2]
	
\def\alphadeg{90}
	\pgfmathsetmacro{\halfalpha}{\alphadeg/2} 

\def\xmin{-0.5}
	\def\xmax{1.5}
	\def\ymin{-0.8}
	\def\ymax{0.8}
	
\def\R{1.0}
	
\draw[->] (\xmin,0) -- (\xmax,0) node[below right] {$\operatorname{Re}\{\Delta\psi(x,y,t)\}$};
	\draw[->] (0,\ymin) -- (0,\ymax) node[above left] {$\operatorname{Im}\{\Delta\psi(x,y,t)\}$};
	
\path[fill=blue!10]
	(0,0) -- ++(\halfalpha:\R)
	arc[start angle=\halfalpha, end angle=-\halfalpha, radius=\R]
	-- cycle;
	
\draw[thick, blue] (0,0) -- ++(\halfalpha:\R);
	\draw[thick, blue] (0,0) -- ++(-\halfalpha:\R);
	
\draw[thick, blue]
	(0,0) ++(\halfalpha:\R)
	arc[start angle=\halfalpha, end angle=-\halfalpha, radius=\R];
	
\def\rang{0.28}
	\draw (0,0) ++(-\halfalpha:\rang)
	arc[start angle=-\halfalpha, end angle=\halfalpha, radius=\rang];
	\node at (0.34,0) {$\alpha$};
	
\def\rone{0.90}   \def\thetaone{30}
	\def\rtwo{0.70}   \def\thetatwo{-20}
	\def\rthree{0.96} \def\thetathree{8}
	
	\filldraw[red] ({\rone*cos(\thetaone)},{\rone*sin(\thetaone)}) circle (0.02)
	node[right] {$\Delta\psi(x_1,y_1,t)$};
	
	\filldraw[red] ({\rtwo*cos(\thetatwo)},{\rtwo*sin(\thetatwo)}) circle (0.02)
	node[right] {$\Delta\psi(x_2,y_2,t)$};
	
	\filldraw[red] ({\rthree*cos(\thetathree)},{\rthree*sin(\thetathree)}) circle (0.02)
	node[right] {$\Delta\psi(x_3,y_3,t)$};
	
\end{tikzpicture}

	\caption{Circular-sector constraint for the inter-frame factor $\Delta\psi(x,y,t)$.
	Admissible values satisfy $|\Delta\phi(x,y,t)|\le \alpha/2$ and $|\Delta\psi(x,y,t)|\le 1$.}
\label{fig:circular-sector-in-complex-plane}
\end{figure}
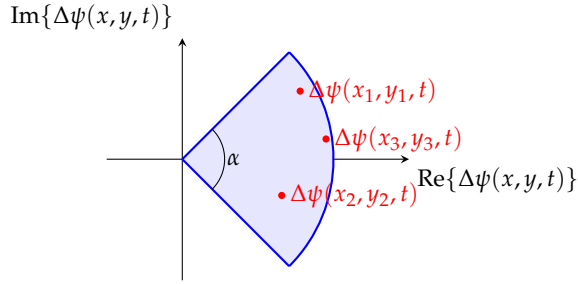 
The physical prior we impose is a bounded rate of change of the OPL,
\begin{equation}
	\left|\frac{\partial f(x,y,t)}{\partial t}\right|
	\approx \frac{|\Delta f(x,y,t)|}{\Delta t}
	\le \varepsilon,
	\label{eq:bounded_dfdt}
\end{equation}
which through \eqref{eq:thin_object} bounds the inter-frame phase increment,
\begin{equation}
	\left|\Delta\phi(x,y,t)\right|
	= \left|\frac{2\pi}{\lambda}\Delta f(x,y,t)\right|
	\le \frac{2\pi}{\lambda}\varepsilon\Delta t
	= \frac{\alpha}{2},
	\qquad 0 \le \alpha < 2\pi.
	\label{eq:phase_increment_bound}
\end{equation}
The increment can be of either sign at different spatial pixels. A second bound comes from passivity: absorption corresponds to a nonnegative imaginary part of the refractive index, so the sample cannot amplify the field at the object plane, giving $|\Delta\psi(x,y,t)|\le 1$. Together, \eqref{eq:phase_increment_bound} and the passive amplitude condition
confine each pixel of $\Delta\psi$ to the circular sector
$\mathcal{C}_\alpha$, defined by
\begin{equation}
		\mathcal{C}_\alpha
		= \Bigl\{ z\in\mathbb{C}:\; z = re^{i\theta},\;
		0\le r \le r_{\max}, -\frac{\alpha}{2}\le \theta \le \frac{\alpha}{2}
		\Bigr\}.
	\label{eq:cone_def}
\end{equation}
For the inter-frame factor $\Delta\psi$, we use $r_{\max}=1$.
This circular-sector constraint is illustrated in
Fig.~\ref{fig:circular-sector-in-complex-plane}.
A consequence is that the phase trajectory is recovered as the cumulative sum
\begin{equation}
	\phi(x,y,j\Delta t) = \phi(x,y,0) + \sum_{k=0}^{j-1}\Delta\phi(x,y,k\Delta t),
	\label{eq:phase_cumsum}
\end{equation}
which follows from \eqref{eq:psi_from_delta_product}. Because each increment $\Delta\phi$ is bounded to $[-\alpha/2,\alpha/2] \subset (-\pi,\pi)$ by \eqref{eq:phase_increment_bound}, each term in the cumulative sum is unambiguous modulo $2\pi$, so the full trajectory is recovered without phase unwrapping. This holds even when the accumulated phase $\phi(x,y,j\Delta t)$ spans many $2\pi$ wraps: the wraps are resolved automatically by sequential integration of small, bounded increments.

The half-angle $\alpha/2$ has a direct physical interpretation: it is the maximum inter-frame phase change allowed at each pixel, determined by the rate of change of the sample dynamics and the frame rate of the diffraction movie. A faster frame rate, or slower sample dynamics, gives a tighter bound and therefore a smaller sector; $\alpha = 2\pi$ recovers the unconstrained case.

\subsection{Feasibility problem and projection operators}\label{sec:method-projectors}

With the multiplicative parameterization and the circular-sector constraint established, we now formulate the reconstruction as a feasibility problem and describe the two projection operators that enforce the constraints.

The reconstruction problem is: given the diffraction movie $I_{\mathrm{meas}}(x',y',t)$, find the sequence of inter-frame factors $\Delta\psi(x,y, t)$ that simultaneously belong to (i) the circular-sector set $\mathcal{C}_\alpha$ of \eqref{eq:cone_def}, and (ii) the set of inter-frame factors whose induced diffraction intensities are consistent with the measurements. We refer to the first as the object-domain constraint and the second as the measurement-domain constraint.

\paragraph{Object-domain projector.} The object-domain projector $\Pi_o$ is the pixelwise Euclidean projection onto the circular sector:
\begin{equation}
	\Pi_o:\ \Delta\psi \mapsto \Delta\psi_o,
	\qquad
	\Delta\psi_o(x,y,t)
	=
	\argmin_{z\in\mathcal{C}_\alpha}
	\bigl|z-\Delta\psi(x,y,t)\bigr|^2.
	\label{eq:Pi_o_def}
\end{equation}
For $\alpha \le \pi$, the set $\mathcal{C}_\alpha$ is convex and the projection is unique. Geometrically, a pixel already inside the sector is left unchanged; a pixel outside is mapped to the nearest boundary point, with the amplitude clamped to $[0, r_{\text{max}}]$. The explicit closed-form expression, broken into four cases according to where the pixel falls relative to the sector boundaries, is given in Appendix~\ref{sec:appendix-Pio}.

The circular-sector constraint generalizes the classical nonnegativity constraint (the degenerate case $\alpha = 0$ and $r_{\max} = \infty$) and is related to the histogram constraints used in X-ray crystallography \cite{Elser2003PhaseRetrievalIteratedProjections}, with the distinction that histogram constraints prescribe the full value distribution on the real line whereas the circular-sector constraint prescribes only the support of the pixel-value distribution in the complex plane.

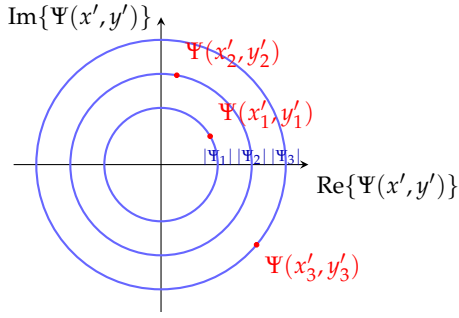
\begin{figure}[t]
	\centering
\begin{tikzpicture}[>=stealth, scale=1.5]
	
\def\rone{0.5}
	\def\rtwo{0.8}
	\def\rthree{1.1}
	
\def\phiOneDeg{30}
	\def\phiTwoDeg{80}
	\def\phiThreeDeg{-40}
	
\pgfmathsetmacro{\phiOneRad}{\phiOneDeg*pi/180}
	\pgfmathsetmacro{\phiTwoRad}{\phiTwoDeg*pi/180}
	\pgfmathsetmacro{\phiThreeRad}{\phiThreeDeg*pi/180}
	\pgfmathsetmacro{\phiOneNorm}{\phiOneRad/pi}
	\pgfmathsetmacro{\phiTwoNorm}{\phiTwoRad/pi}
	\pgfmathsetmacro{\phiThreeNorm}{\phiThreeRad/pi}
	
\draw[->] (-1.3,0) -- (1.3,0) node[below right] {$\operatorname{Re}\{\Psi(x',y')\}$};
	\draw[->] (0,-1.3) -- (0,1.3) node[left] {$\operatorname{Im}\{\Psi(x',y')\}$};
	
\draw[blue!60, thick] (0,0) circle (\rone);
	\draw[blue!60, thick] (0,0) circle (\rtwo);
	\draw[blue!60, thick] (0,0) circle (\rthree);
	
\node[blue!70!black] at (\rone,0.07) {\scriptsize $|\Psi_1|$};
	\node[blue!70!black] at (\rtwo,0.07) {\scriptsize $|\Psi_2|$};
	\node[blue!70!black] at (\rthree,0.07) {\scriptsize $|\Psi_3|$};
	
\filldraw[red] ({\rone*cos(\phiOneDeg)},{\rone*sin(\phiOneDeg)}) circle (0.02)
	node[above right] {$\Psi(x'_1,y'_1)$};
	\filldraw[red] ({\rtwo*cos(\phiTwoDeg)},{\rtwo*sin(\phiTwoDeg)}) circle (0.02)
	node[above right] {$\Psi(x'_2,y'_2)$};
	\filldraw[red] ({\rthree*cos(\phiThreeDeg)},{\rthree*sin(\phiThreeDeg)}) circle (0.02)
	node[below right] {$\Psi(x'_3,y'_3)$};

\end{tikzpicture}

	\caption{Modulus constraint in the complex plane at the detector: for each detector pixel, $\Psi(x',y',t)$ is constrained to a circle of radius $I_{\text{meas}}(x',y',t)$.}
\label{fig:torus-in-complex-plane}
\end{figure}

\paragraph{Measurement-domain projector.} The measurement-domain projector $\Pi_m$ enforces consistency with the recorded intensity. It is built on the Fresnel propagator $\mathcal{P}$ that propagates the object-plane field to the detector plane:
\begin{equation}
	\begin{aligned}
		\Psi(x',y')
		& = \mathcal{P}\{\psi\}(x',y') 
		= \frac{\exp(i2\pi z/\lambda)}{i\lambda z} \times \\
		& \iint \psi(x,y)\,
		\exp\!\left[
		\frac{i\pi}{\lambda z}
		\bigl((x-x')^2+(y-y')^2\bigr)
		\right]
		\,dx\,dy .
	\end{aligned}
	\label{eq:fresnel_integral}
\end{equation}
with inverse $\mathcal{P}^{-1} = \mathcal{P}_{-z}$, propagation by $-z$, which follows from the unitarity of $\mathcal{P}$. The projector $\Pi_m$ acts in three steps: propagate the current field estimate to the detector,
\begin{equation}
	\Psi(x',y',t) = \mathcal{P}\{\psi(\cdot,\cdot,t)\}(x',y')
	= |\Psi(x',y',t)|\,\exp\!\bigl(i\,\Phi(x',y',t)\bigr),
	\label{eq:fwd_prop}
\end{equation}
replace the computed amplitude with the measured one while retaining the computed phase,
\begin{equation}
	\Psi_m(x',y',t)
	=
	\sqrt{I(x',y',t)}\,\exp\!\bigl(i\,\Phi(x',y',t)\bigr),
	\label{eq:mag_replace}
\end{equation}
and propagate back to the object plane,
\begin{equation}
	\psi_m(x,y,t)
	=
	\mathcal{P}^{-1}\!\left\{\Psi_m(\cdot,\cdot,t)\right\}(x,y).
	\label{eq:bwd_prop}
\end{equation}
The three steps together define $\Pi_m: \psi \mapsto \psi_m$. This is the classical Gerchberg--Saxton--Fienup magnitude projection \cite{GerchbergSaxton1972, Fienup1978ReconstructionModulusFT, Fienup1982PRComparison}, applied here to each time frame. The magnitude replacement \eqref{eq:mag_replace} is the Euclidean projection of $\Psi$ onto the set of fields with modulus $\sqrt{I(\cdot,\cdot,t)}$ at each detector pixel, illustrated in Fig.~\ref{fig:torus-in-complex-plane}.

\paragraph{Extension to the inter-frame factor.} Because the unknowns of the reconstruction are the inter-frame factors $\Delta\psi$ rather than the full field $\psi$, we extend $\Pi_m$ to an operator that acts on $\Delta\psi$ directly. Two variants arise from the multiplicative coupling between consecutive frames,
\begin{equation}
	\psi(x,y,j\Delta t)
	=
	\psi(x,y,(j-1)\Delta t)\cdot\Delta\psi(x,y,(j-1)\Delta t).
	\label{eq:psi_j_from_delta}
\end{equation}
In the forward variant $\Pi^{\Delta}_{m,\mathrm{fwd}}$, the previous field $\psi(x,y,(j-1)\Delta t)$ is held fixed; the operator finds the minimal change to $\Delta\psi(x,y,(j-1)\Delta t)$ such that the resulting next frame $\psi(x,y,j\Delta t)$ is consistent with the measurement $I(x',y',j\Delta t)$. In the backward variant $\Pi^{\Delta}_{m,\mathrm{bwd}}$, the next field is held fixed instead, and the operator enforces consistency with the measurement at frame $j-1$. The explicit closed-form expressions for both variants, including a Tikhonov regularization that stabilizes the division by a small denominator when the previous or next field has near-zero amplitude, are given in Appendix~\ref{sec:appendix-Pim}.

\subsection{Reconstruction algorithm}\label{sec:algo}

The reconstruction proceeds in two stages: first the probe (the static initial field $\psi(x,y,0)$), then the inter-frame factors $\Delta\psi(x,y,t)$ that evolve it through time. Both stages share the same alternating projection scheme, with reflections at the intermediate step to overcome the stagnation typical of plain alternating projections when one of the constraint sets (here, the measurement-modulus set) is non-convex. We use blocks of relaxed-reflect-reflect (RRR) and relaxed averaged alternating reflections (RAAR) iterations; the explicit maps and their fixed-point properties are given in Appendix~\ref{sec:appendix-iter}.

Stage 1 reconstructs the probe $\psi(x,y,0)$ from the first measured frame $I(x',y',0)$. The two constraint sets are the circular-sector set $\mathcal{C}_\alpha$ applied to $\psi(x,y,0)$, and the measurement-modulus set for $I(x',y',0)$. Because the amplitude of $\psi(x,y,0)$ is set by the illumination intensity rather than the thin-object bound, the radius cap $r_{\max}$ for the probe is treated as a free parameter $r_0$: it is decreased from a large initial value until the two constraint sets become incompatible and the iterations diverge, and set to the smallest value that still yields stable convergence. The resulting $\psi(x,y,0)$ is held fixed for Stage 2.

Stage 2 reconstructs the inter-frame factors $\Delta\psi(x,y,t)$ for $t = \Delta t, \dots, J\Delta t$ with $\psi(x,y,0)$ fixed. Rather than solving for all $J$ factors simultaneously, Stage 2 processes one frame at a time in a forward pass followed by a backward pass.

In the forward pass, frames are processed in the order $j = 1, 2, \dots, J$. At step $j$, the previous field $\psi(x,y,(j-1)\Delta t)$ is held fixed at the value produced by the preceding step, and RRR/RAAR iterations are run on $\Delta\psi(x,y,(j-1)\Delta t)$ with $\Pi_o$ as the object-domain projector and $\Pi^{\Delta}_{m,\mathrm{fwd}}$ as the measurement-domain projector. Once the iterations converge, the forward estimate $\Delta\psi^{(\mathrm{fwd})}(x,y,(j-1)\Delta t)$ is stored, and the next field is advanced:
\begin{equation}
	\psi(x,y,j\Delta t)
	\leftarrow
	\psi(x,y,(j-1)\Delta t)\cdot
	\Delta\psi^{(\mathrm{fwd})}(x,y,(j-1)\Delta t).
	\label{eq:fwd_advance}
\end{equation}
After all $J$ steps, the forward pass produces the complete field trajectory $\psi(x,y,0), \psi(x,y,\Delta t), \dots, \psi(x,y,J\Delta t)$.

The backward pass then processes frames in the reverse order $j = J, \dots, 1$. At step $j$, the next field $\psi(x,y,j\Delta t)$ is held fixed at the value from the backward chain, and RRR/RAAR iterations are run on $\Delta\psi(x,y,(j-1)\Delta t)$ with $\Pi^{\Delta}_{m,\mathrm{bwd}}$ as the measurement-domain projector. The backward estimate $\Delta\psi^{(\mathrm{bwd})}(x,y,(j-1)\Delta t)$ is stored, and the chain is propagated backward as
\begin{equation}
\psi(x,y,(j-1)\Delta t)
\leftarrow
\frac{\psi(x,y,j\Delta t)\,
	\Delta\psi^{(\mathrm{bwd})*}(x,y,(j-1)\Delta t) + \epsilon^2}
{|\Delta\psi^{(\mathrm{bwd})}(x,y,(j-1)\Delta t)|^2 + \epsilon^2}.
	\label{eq:bwd_advance}
\end{equation}

The forward and backward estimates are then combined framewise. The amplitude is taken as the geometric mean of the two estimates,
\begin{equation}
	\bigl|\Delta\psi^{(\mathrm{avg})}\bigr|
	=
	\sqrt{
		\bigl|\Delta\psi^{(\mathrm{fwd})}\bigr|
		\cdot
		\bigl|\Delta\psi^{(\mathrm{bwd})}\bigr|
	},
	\label{eq:avg_amp}
\end{equation}
and the phase is the circular mean,
\begin{equation}
	\Delta\phi^{(\mathrm{avg})}
	=
	\frac{
		\Delta\phi^{(\mathrm{fwd})}
		+
		\Delta\phi^{(\mathrm{bwd})}
	}{2}.
	\label{eq:avg_phase}
\end{equation}
The combined estimate $\Delta\psi^{(\mathrm{avg})}$ is used to form the full phase trajectory via \eqref{eq:phase_cumsum}. Combining forward and backward passes in this way balances the directional bias that either pass alone would introduce: the forward pass concentrates error at late frames, as small inter-frame errors accumulate through forward integration, while the backward pass concentrates error at early frames.

In both stages we alternate blocks of RRR and RAAR iterations, with 25 iterations per block and four blocks in total. We used $\beta = 0.8$ throughout, and $\alpha = \pi$ for the circular-sector constraint in Stage 2 for all experiments reported here.
 	\section{Experimental demonstrations}\label{sec:result}

We demonstrate the framework on three experiments of increasing difficulty. We first validate the approach using a spatial light modulator (SLM) as a calibrated dynamic sample, where the ground-truth phase trajectory is known by construction. Two SLM experiments are presented: a reaction--diffusion dynamics that tests spatially expanding phase patterns, and a pure-growth dynamics that tests the recovery of accumulated phase well beyond $2\pi$. We then apply the same framework, without modification, to in-situ monitoring of a photo-polymer 3D printing process, where the phase trajectory is not known a priori and the goal is quantitative characterization of the spatiotemporal phase induced by photopolymerization.

\begin{figure*}[t]
	\centering
	
	\newcommand{\snapH}{2.35cm}      \newcommand{\snapW}{0.13\textwidth} \newcommand{\cbW}{0.01\textwidth}   \newcommand{\gap}{0.5pt}          

\begin{subfigure}[t]{\snapW}
		\centering
		\includegraphics[height=\snapH]{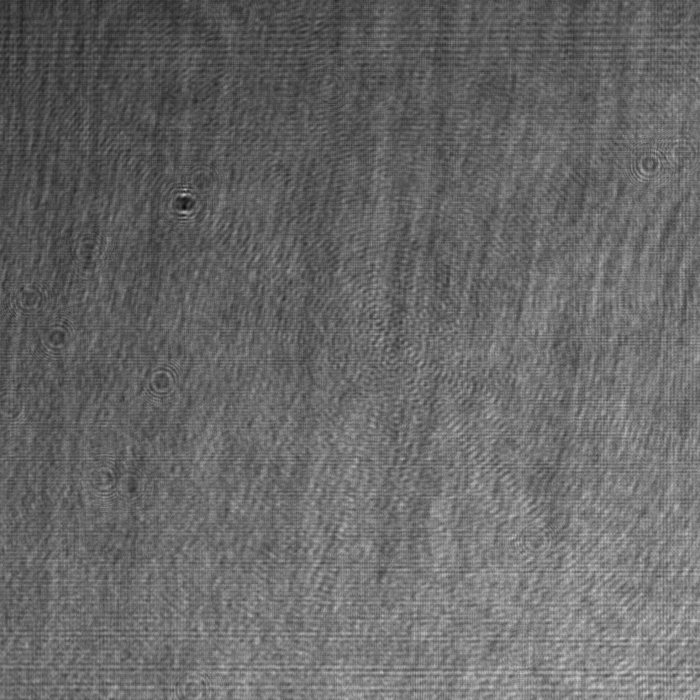}
		\caption{$t=0$}
		\label{fig:I_xy0}
	\end{subfigure}\hspace{\gap}
	\begin{subfigure}[t]{\snapW}
		\centering
		\includegraphics[height=\snapH]{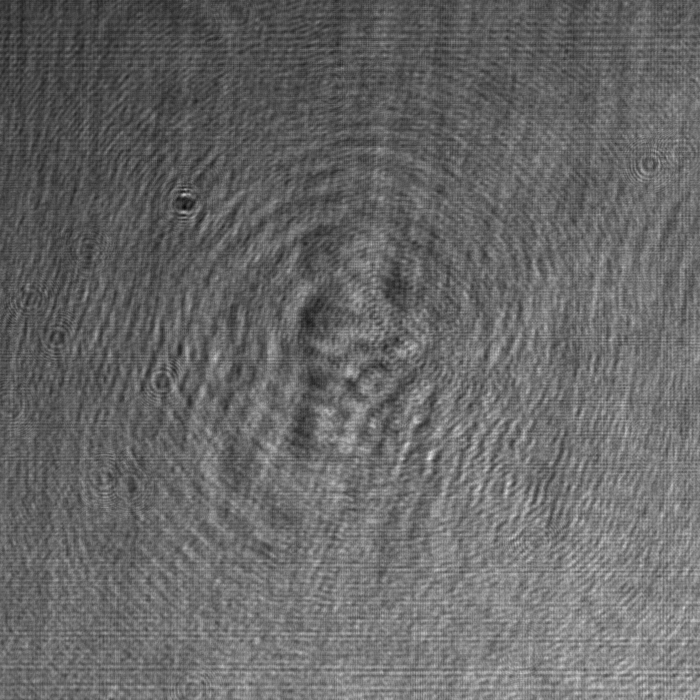}
		\caption{$t=1\Delta t$}
	\end{subfigure}\hspace{\gap}
	\begin{subfigure}[t]{\snapW}
		\centering
		\includegraphics[height=\snapH]{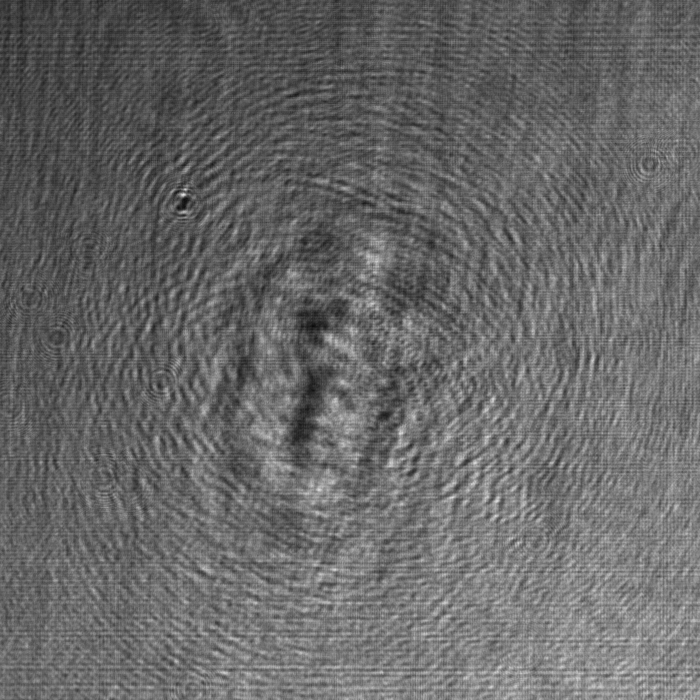}
		\caption{$t=7\Delta t$}
	\end{subfigure}\hspace{\gap}
	\begin{subfigure}[t]{\snapW}
		\centering
		\includegraphics[height=\snapH]{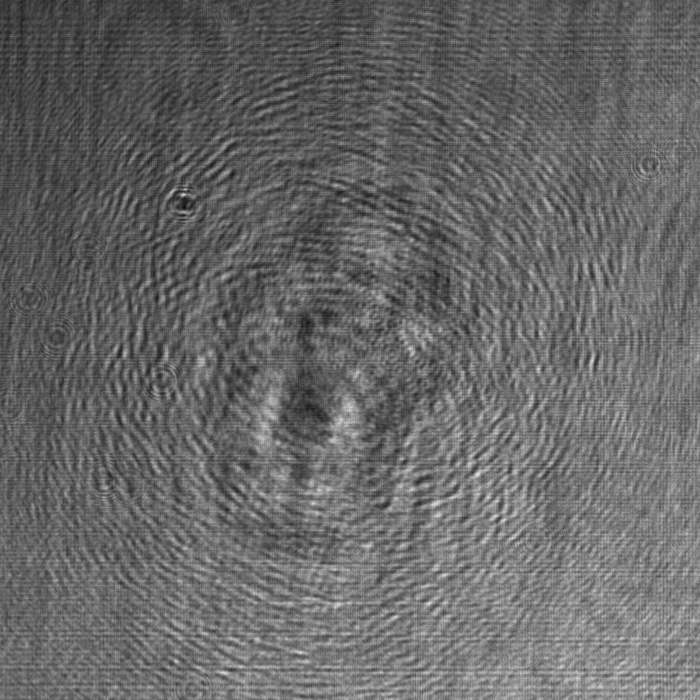}
		\caption{$t=12\Delta t$}
	\end{subfigure}\hspace{\gap}
	\begin{subfigure}[t]{\snapW}
		\centering
		\includegraphics[height=\snapH]{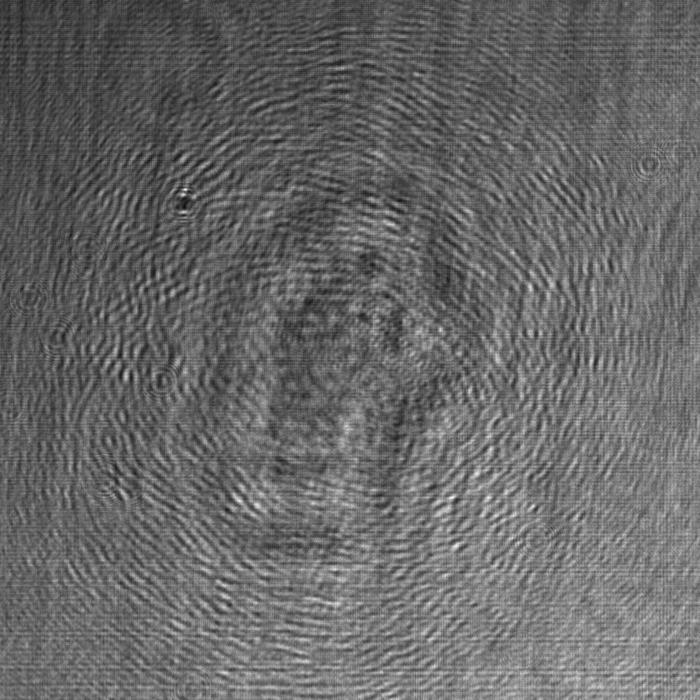}
		\caption{$t=18\Delta t$}
	\end{subfigure}\hspace{\gap}
	\begin{subfigure}[t]{\cbW}
		\centering
		\captionsetup{labelformat=empty} \includegraphics[height=\snapH]{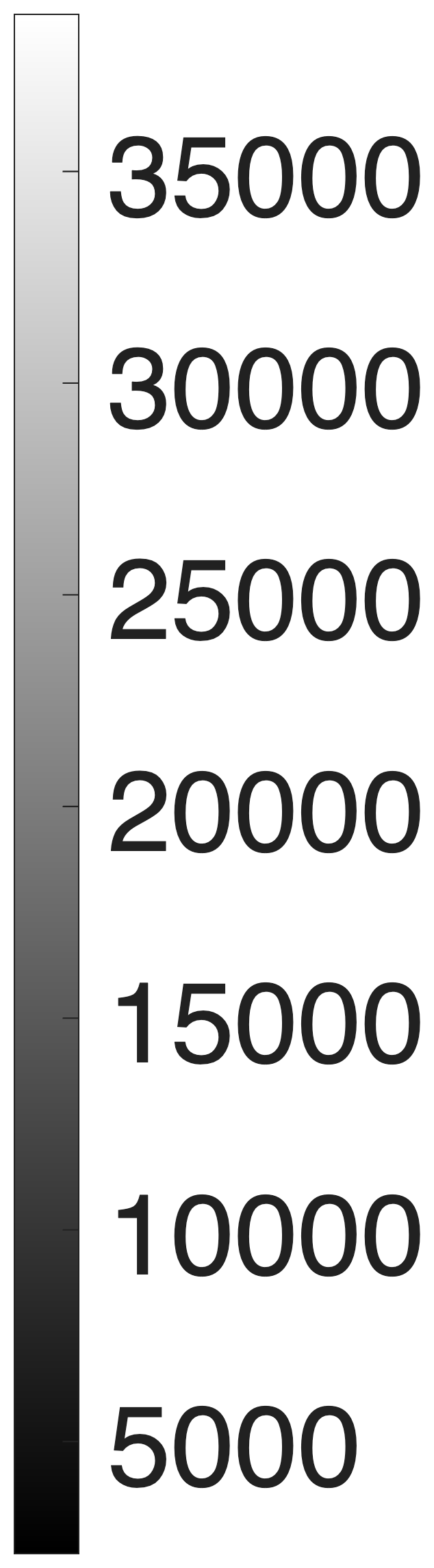}
\end{subfigure}
	
\begin{subfigure}[t]{\snapW}
		\centering
		\includegraphics[height=\snapH]{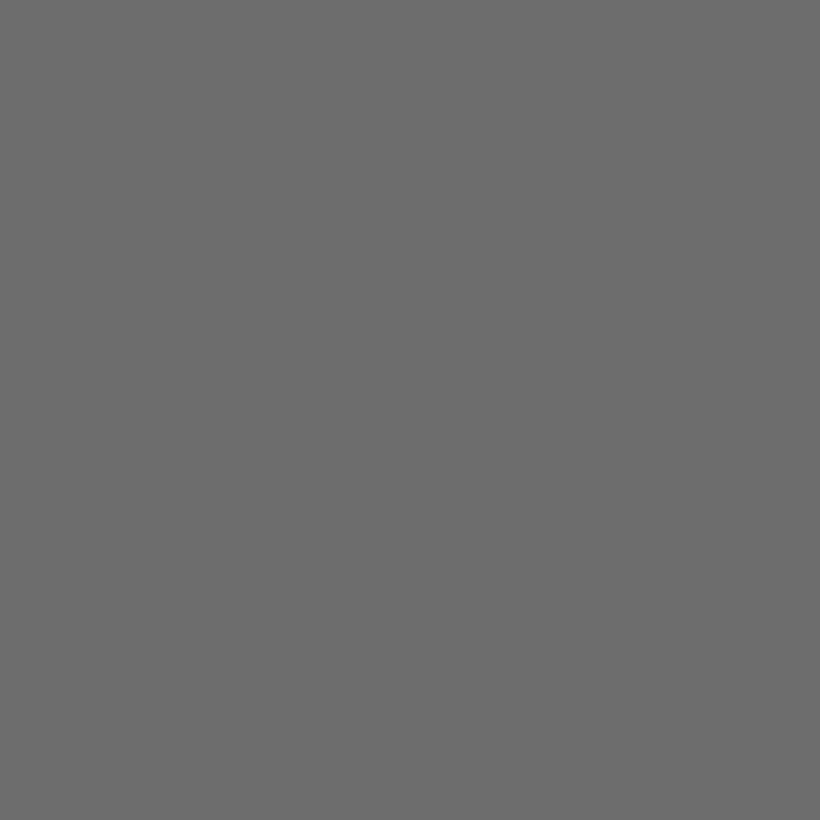}
		\caption{$t=0$}
	\end{subfigure}\hspace{\gap}
	\begin{subfigure}[t]{\snapW}
		\centering
		\includegraphics[height=\snapH]{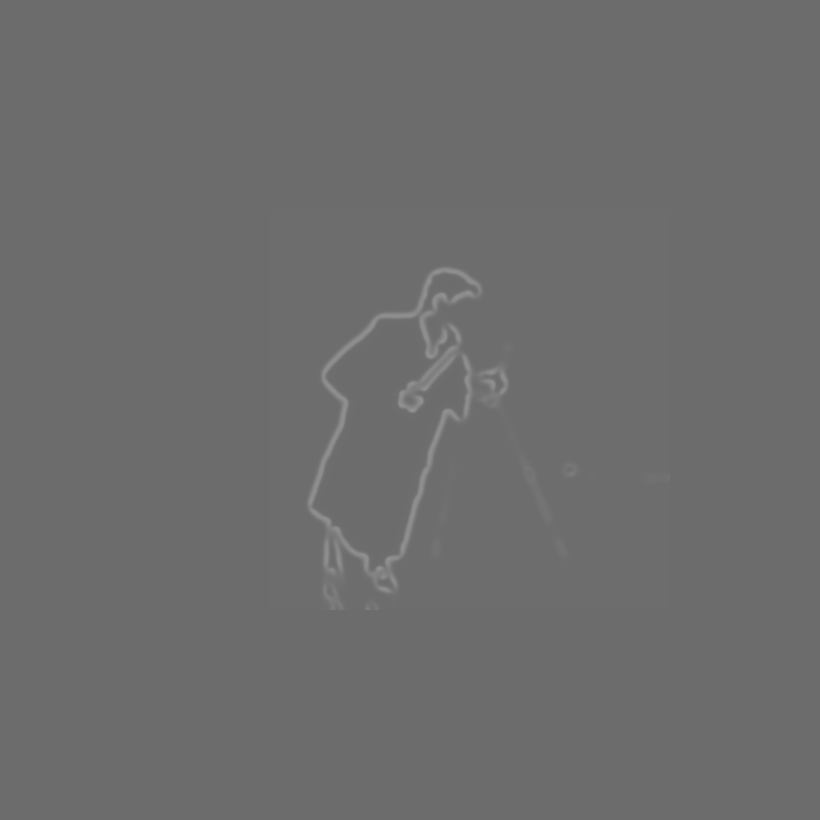}
		\caption{$t=1\Delta t$}
	\end{subfigure}\hspace{\gap}
	\begin{subfigure}[t]{\snapW}
		\centering
		\includegraphics[height=\snapH]{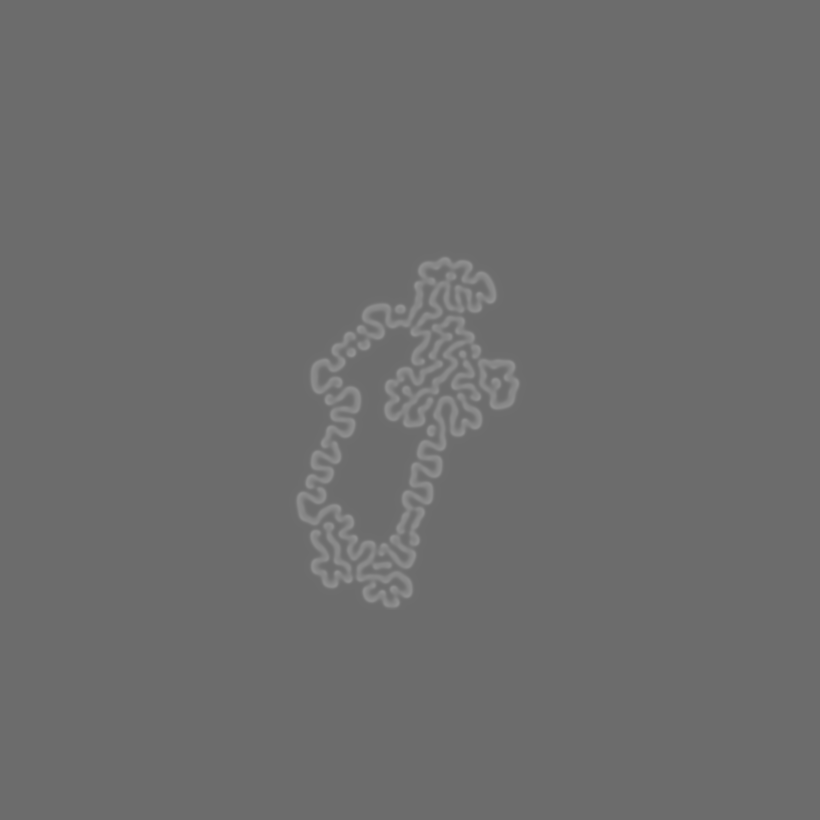}
		\caption{$t=7\Delta t$}
	\end{subfigure}\hspace{\gap}
	\begin{subfigure}[t]{\snapW}
		\centering
		\includegraphics[height=\snapH]{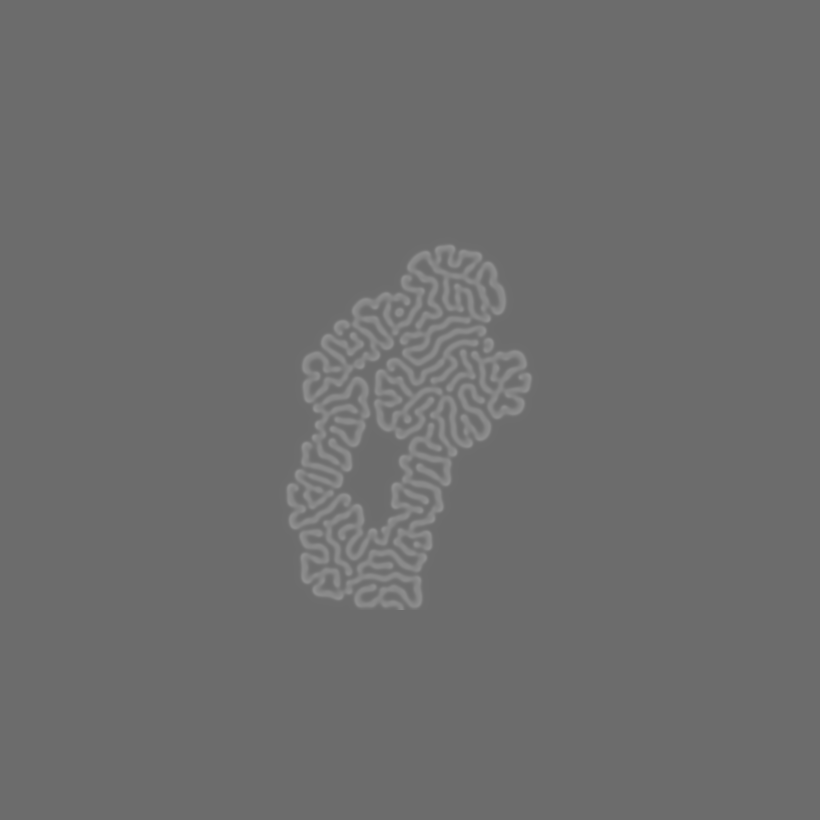}
		\caption{$t=12\Delta t$}
	\end{subfigure}\hspace{\gap}
	\begin{subfigure}[t]{\snapW}
		\centering
		\includegraphics[height=\snapH]{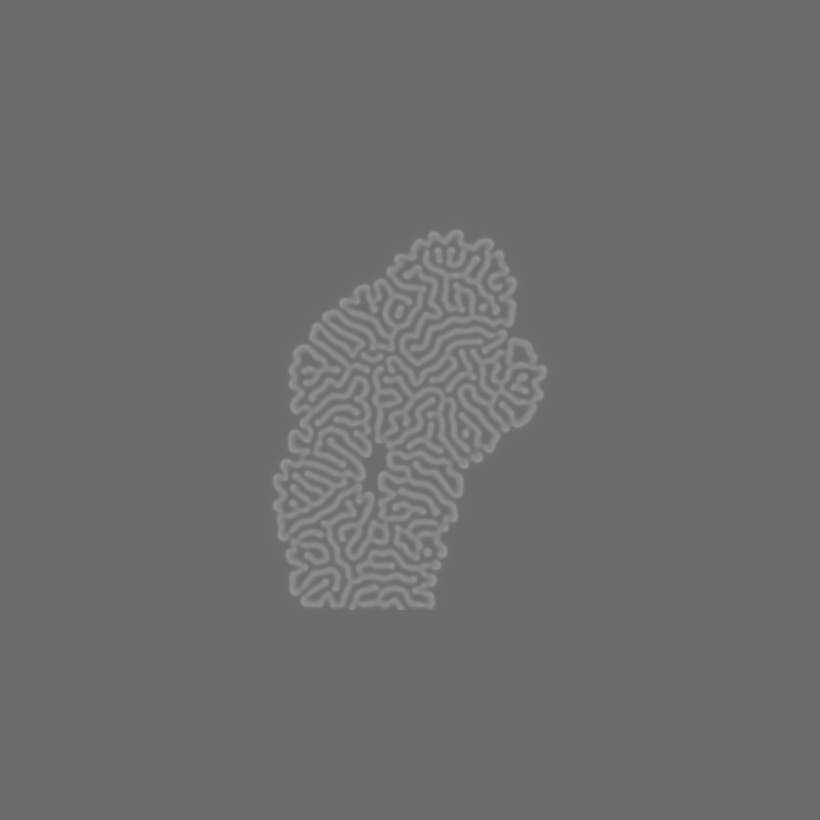}
		\caption{$t=18\Delta t$}
	\end{subfigure}\hspace{\gap}
	\begin{subfigure}[t]{\cbW}
		\centering
		\captionsetup{labelformat=empty} \includegraphics[height=\snapH]{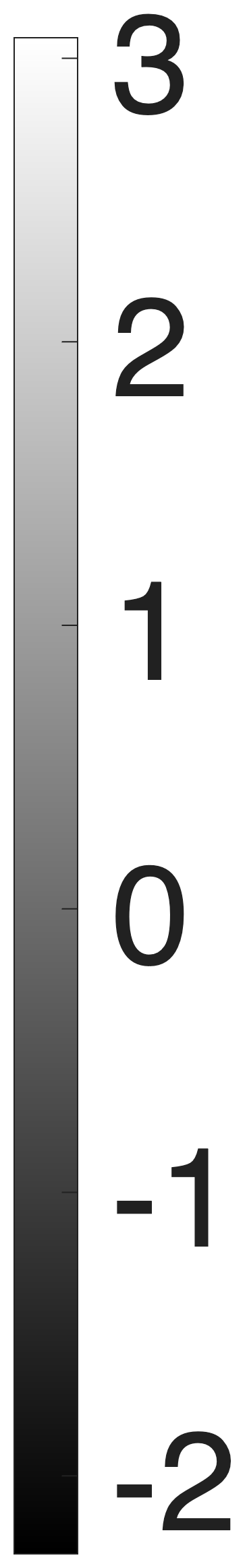}
\end{subfigure}

\begin{subfigure}[t]{\snapW}
		\centering
\includegraphics[height=\snapH]{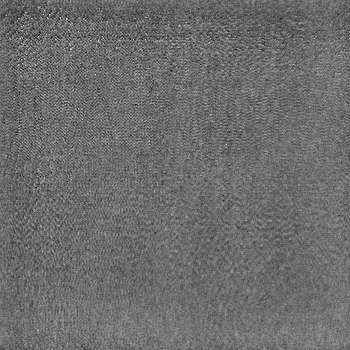}
		\caption{$t=0$}
		\label{fig:psi_xy0}
	\end{subfigure}\hspace{\gap}
	\begin{subfigure}[t]{\snapW}
		\centering
		\includegraphics[height=\snapH]{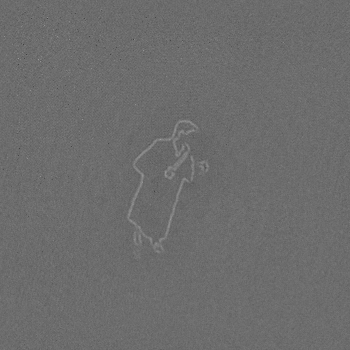}
		\caption{$t=1\Delta t$}
		\label{fig:psi_xyt_t03}
	\end{subfigure}\hspace{\gap}
	\begin{subfigure}[t]{\snapW}
		\centering
		\includegraphics[height=\snapH]{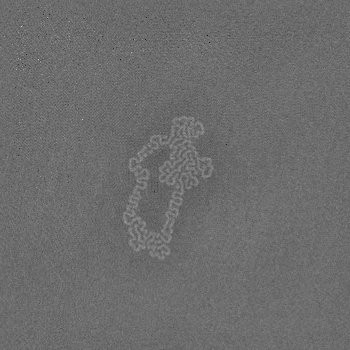}
		\caption{$t=7\Delta t$}
	\end{subfigure}\hspace{\gap}
	\begin{subfigure}[t]{\snapW}
		\centering
		\includegraphics[height=\snapH]{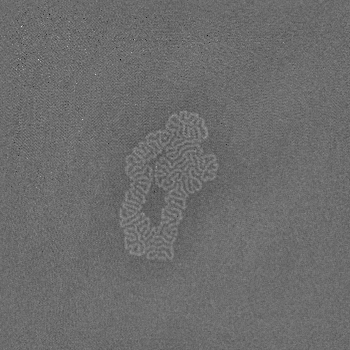}
		\caption{$t=12\Delta t$}
	\end{subfigure}\hspace{\gap}
	\begin{subfigure}[t]{\snapW}
		\centering
		\includegraphics[height=\snapH]{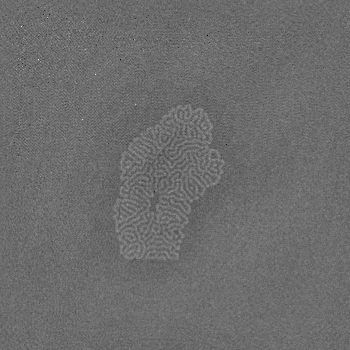}
		\caption{$t=18\Delta t$}
		\label{fig:psi_xyt_t18}
	\end{subfigure}\hspace{\gap}
	\begin{subfigure}[t]{\cbW}
		\centering
		\captionsetup{labelformat=empty} \includegraphics[height=\snapH]{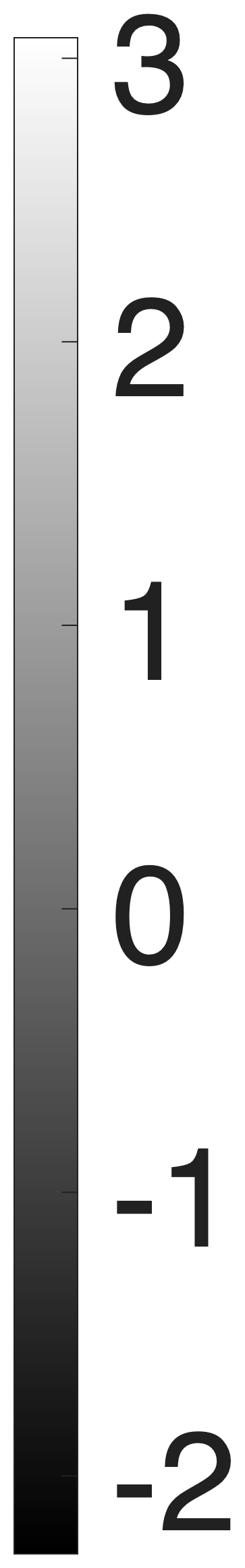}
\end{subfigure}
	
\begin{subfigure}[t]{\snapW}
		\centering
		\includegraphics[height=\snapH]{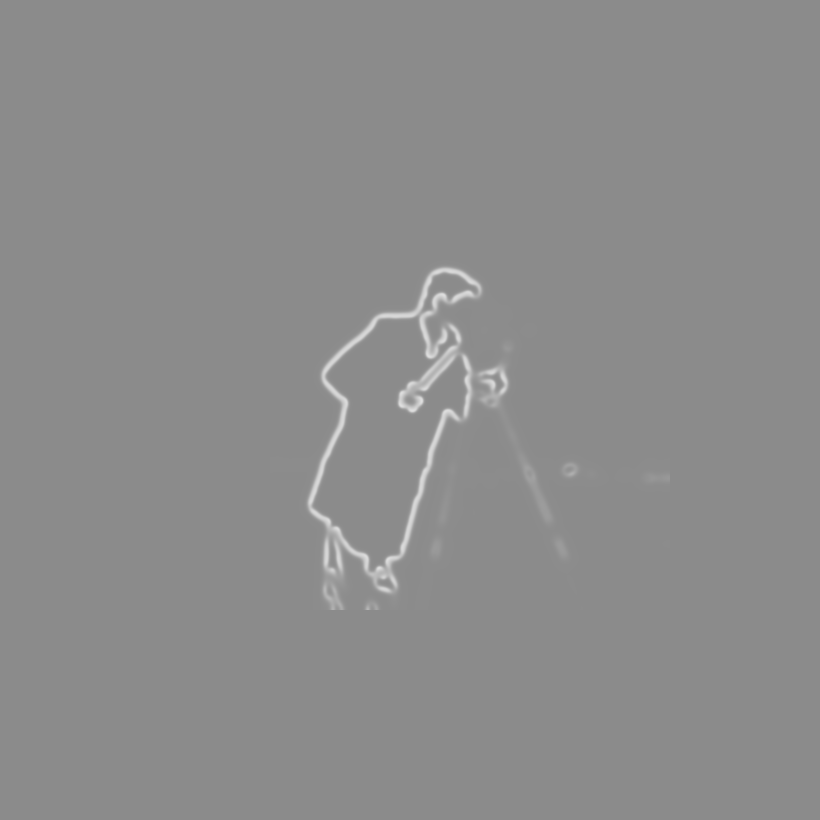}
		\caption{$t=0\Delta t$}
	\end{subfigure}\hspace{\gap}
	\begin{subfigure}[t]{\snapW}
		\centering
		\includegraphics[height=\snapH]{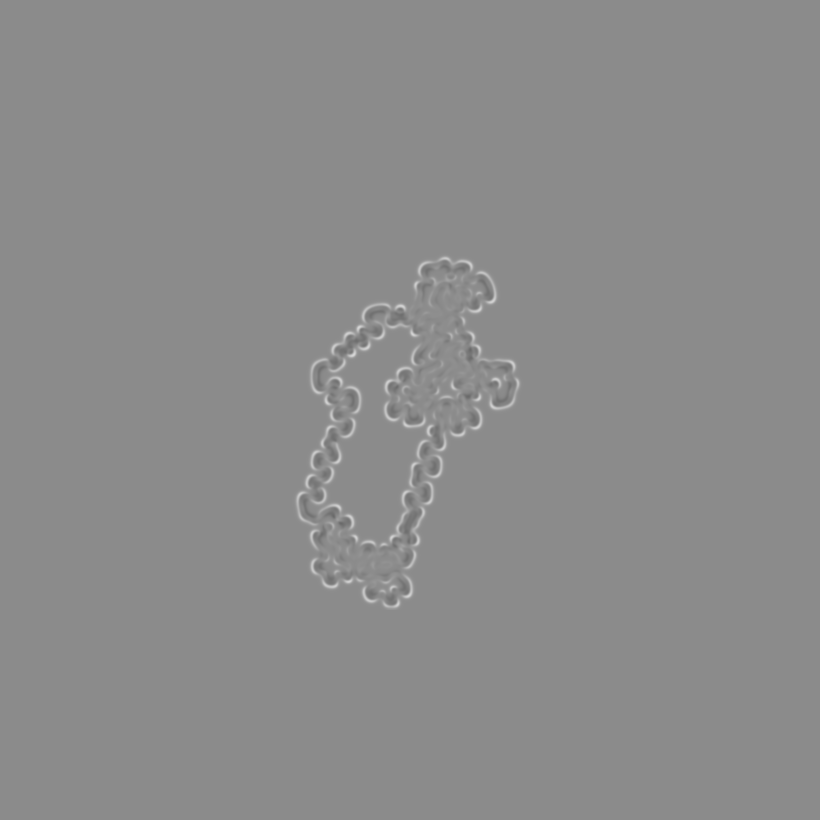}
		\caption{$t=6\Delta t$}
	\end{subfigure}\hspace{\gap}
	\begin{subfigure}[t]{\snapW}
		\centering
		\includegraphics[height=\snapH]{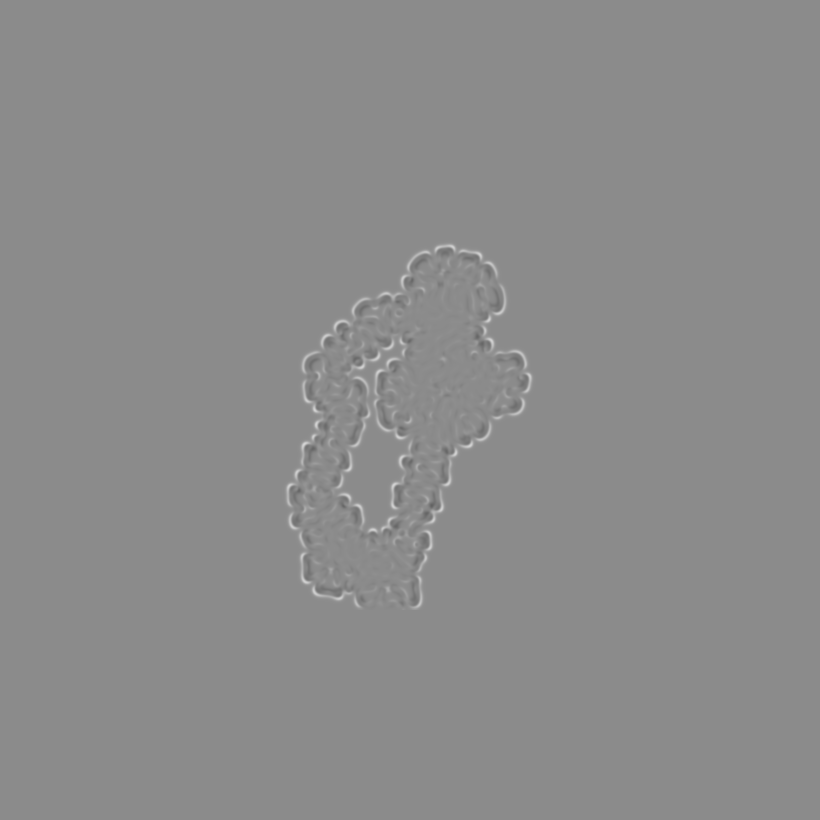}
		\caption{$t=11\Delta t$}
	\end{subfigure}\hspace{\gap}
	\begin{subfigure}[t]{\snapW}
		\centering
		\includegraphics[height=\snapH]{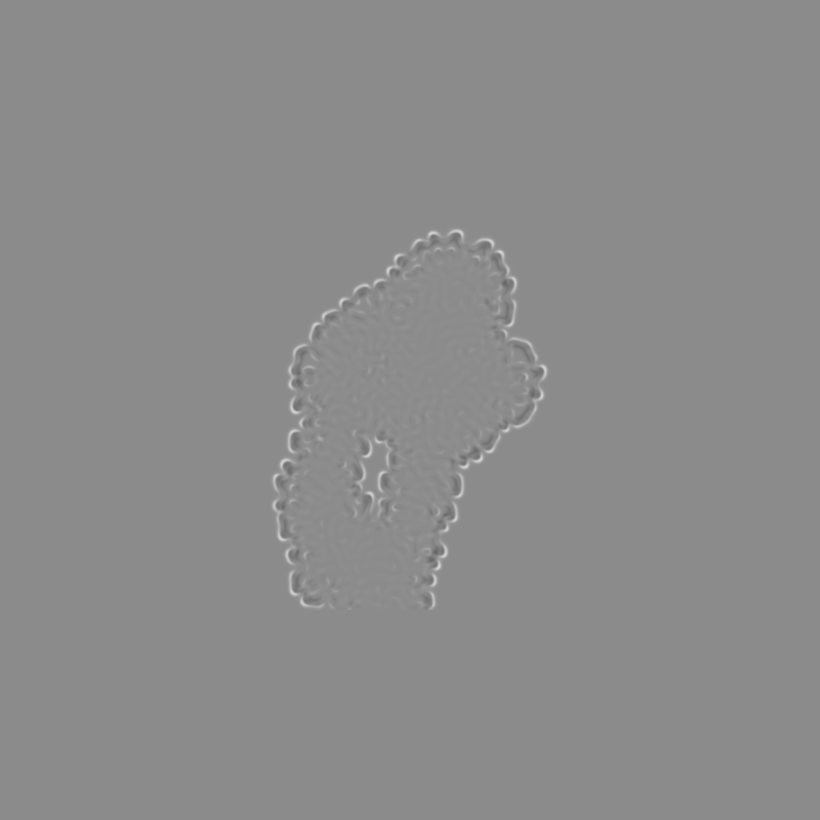}
		\caption{$t=17\Delta t$}
	\end{subfigure}\hspace{\gap}
	\begin{subfigure}[t]{\cbW}
		\centering
		\captionsetup{labelformat=empty} \includegraphics[height=\snapH]{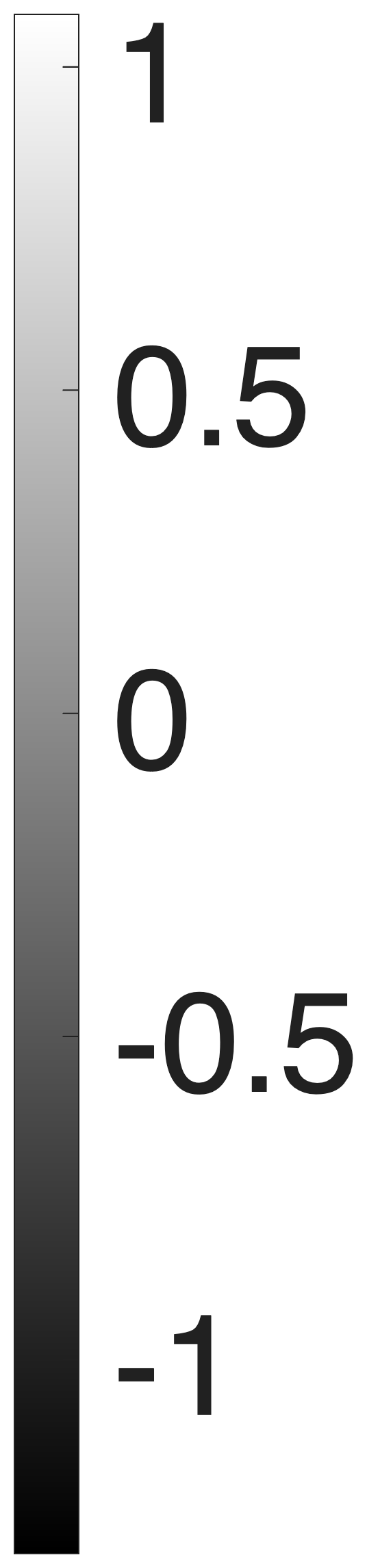}
\end{subfigure}
	
\begin{subfigure}[t]{\snapW}
		\centering
		\includegraphics[height=\snapH]{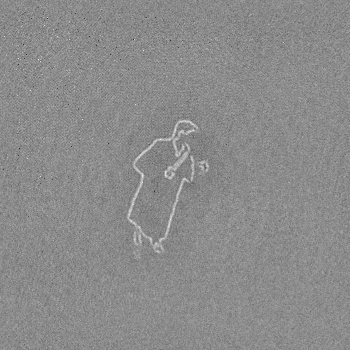}
		\caption{$t=0\Delta t$}
	\end{subfigure}\hspace{\gap}
	\begin{subfigure}[t]{\snapW}
		\centering
		\includegraphics[height=\snapH]{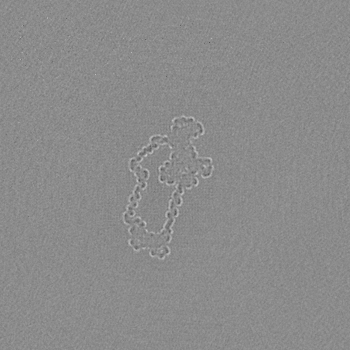}
		\caption{$t=6\Delta t$}
	\end{subfigure}\hspace{\gap}
	\begin{subfigure}[t]{\snapW}
		\centering
		\includegraphics[height=\snapH]{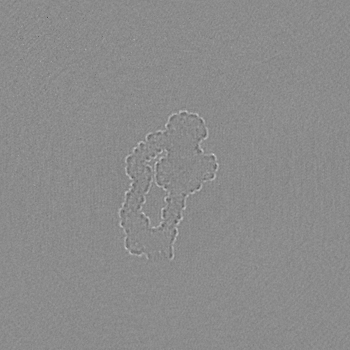}
		\caption{$t=11\Delta t$}
	\end{subfigure}\hspace{\gap}
	\begin{subfigure}[t]{\snapW}
		\centering
		\includegraphics[height=\snapH]{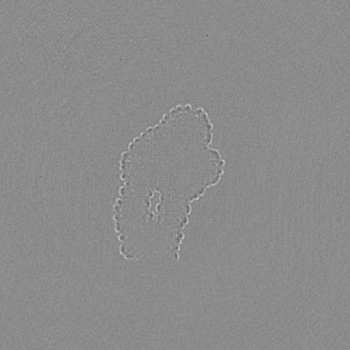}
		\caption{$t=17\Delta t$}
	\end{subfigure}\hspace{\gap}
	\begin{subfigure}[t]{\cbW}
		\centering
		\captionsetup{labelformat=empty} \includegraphics[height=\snapH]{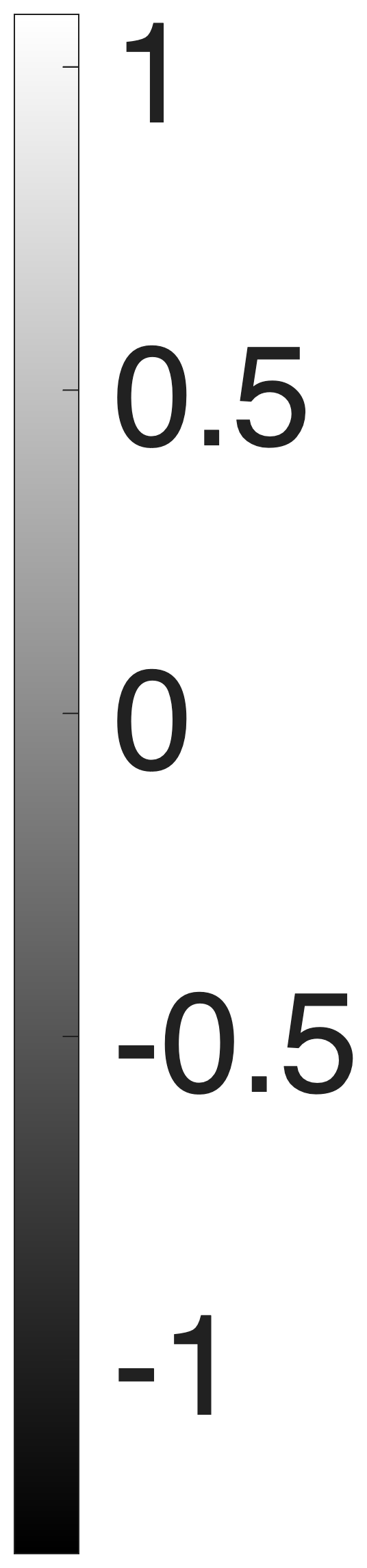}
\end{subfigure}
	
	\caption{
		Dynamic CDI reconstruction from an SLM-driven reaction-diffusion phase trajectory using $\alpha = \pi$.
		Row~1: measurement frames of $I_{\mathrm{meas}}(x',y',t)$.
		Row~2: ground-truth object-plane phase $\phi_{\mathrm{SLM}}(x,y,t)$ displayed on the SLM at a few select times.
		Row~3: reconstructed object-plane phase $\phi(x,y,t)$ at the same times, obtained by integrating the reconstructed inter-frame factors.
		Row~4: ground-truth inter-frame phase $\Delta\phi_{\mathrm{SLM}}(x,y,t)$.
		Row~5: reconstructed inter-frame phase $\Delta\phi(x,y,t)$ at the same times.
		Movies corresponding to rows 1--5 are available online. 
		}
	\label{fig:ReactionDiffusion_result}
\end{figure*} 
\subsection{Reaction-diffusion dynamics}

The first experiment uses a known phase trajectory generated from a Gray--Scott reaction--diffusion model seeded by a resized \textit{cameraman} image, displayed on the SLM as the ground-truth dynamics $\phi_{\mathrm{SLM}}(x,y,t)$. The dynamics consists of a propagating front of a chemical species, which on the SLM appears as a phase pattern that expands spatially over time, with the maximum phase across all frames bounded below $\pi/4$. A uniformly zero-phase frame was prepended to the sequence so that the first recorded diffraction frame in Fig.~\ref{fig:ReactionDiffusion_result}(a) corresponds to the illumination, or probe field. Implementation details, including the Gray--Scott parameters, initialization, and how the simulated state is mapped to a calibrated SLM phase pattern, are provided in Supplementary Material.

The probe field was reconstructed using the Stage~1 procedure of Sec.~\ref{sec:algo}, with the circular-sector parameter set to $\alpha = 1.5\pi$ to accommodate wavefronts with large or discontinuous phase. The phase of the reconstructed probe is shown in Fig.~\ref{fig:ReactionDiffusion_result}(k). With the probe fixed, the inter-frame factors $\Delta\psi(x,y,t)$ were then reconstructed using the Stage~2 forward--backward procedure with $\alpha = \pi$.

Representative frames of the reconstructed inter-frame phase increment $\Delta\phi(x,y,t)$ are shown in Fig.~\ref{fig:ReactionDiffusion_result}(t--w). Compared with the ground-truth $\Delta\phi_{\mathrm{SLM}}$ in Fig.~\ref{fig:ReactionDiffusion_result}(p--s), the reconstructed $\Delta\phi(x,y,t)$ shows good agreement: the propagating front of the chemical species is clearly resolved. The integrated phase trajectory in Fig.~\ref{fig:ReactionDiffusion_result}(l--o) is obtained by summing the inter-frame increments via \eqref{eq:phase_cumsum}, with the initial phase set to zero so that the accumulated phase reflects only the dynamic sample contribution and excludes the static probe phase.

 \begin{figure*}[t]
	\centering
	
	\newcommand{\snapH}{2.35cm}      \newcommand{\snapW}{0.13\textwidth} \newcommand{\cbW}{0.01\textwidth}   \newcommand{\gap}{0.5pt}          

\begin{subfigure}[t]{\snapW}
		\centering
		\includegraphics[height=\snapH]{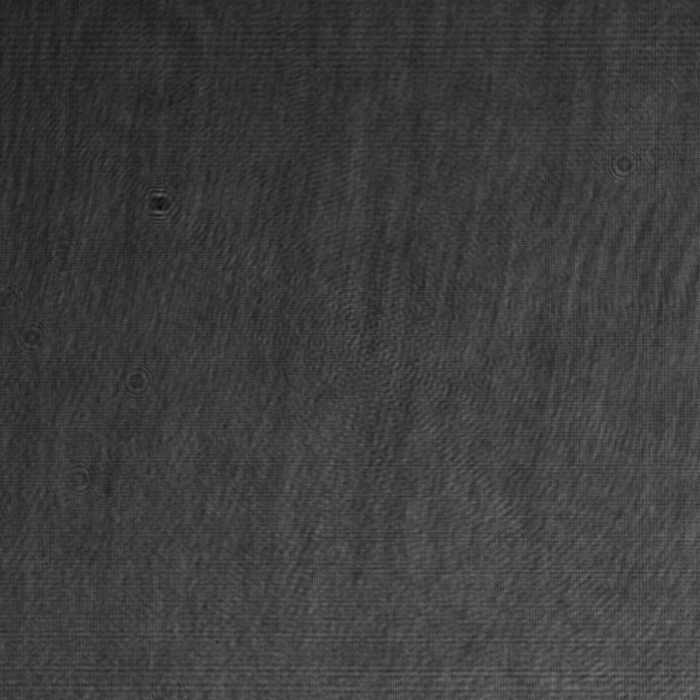}
		\caption{$t=0$}
		\label{fig:EXP-GrowingMIT/I_xy0}
	\end{subfigure}\hspace{\gap}
	\begin{subfigure}[t]{\snapW}
		\centering
		\includegraphics[height=\snapH]{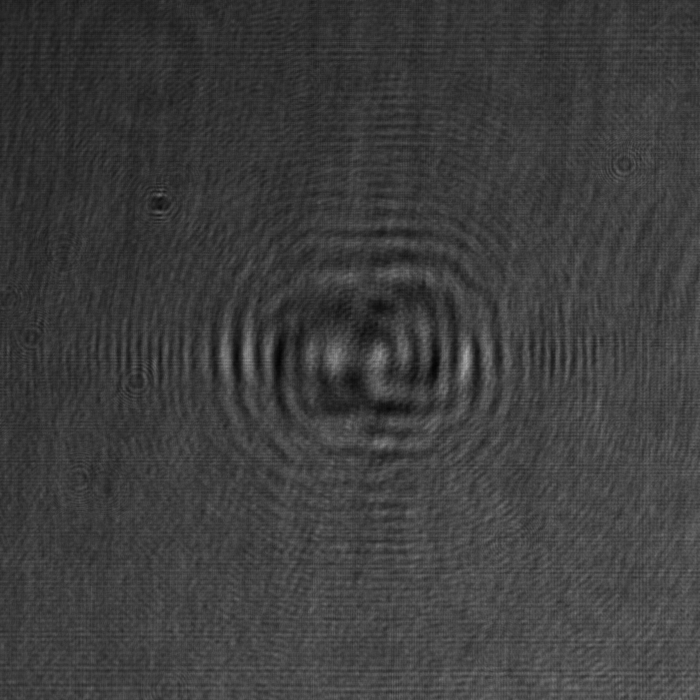}
		\caption{$t=1\Delta t$}
	\end{subfigure}\hspace{\gap}
	\begin{subfigure}[t]{\snapW}
		\centering
		\includegraphics[height=\snapH]{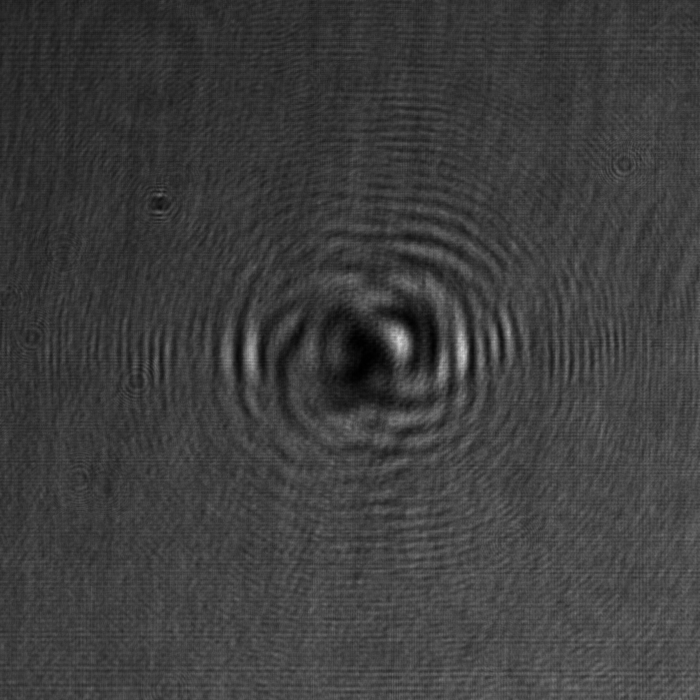}
		\caption{$t=7\Delta t$}
	\end{subfigure}\hspace{\gap}
	\begin{subfigure}[t]{\snapW}
		\centering
		\includegraphics[height=\snapH]{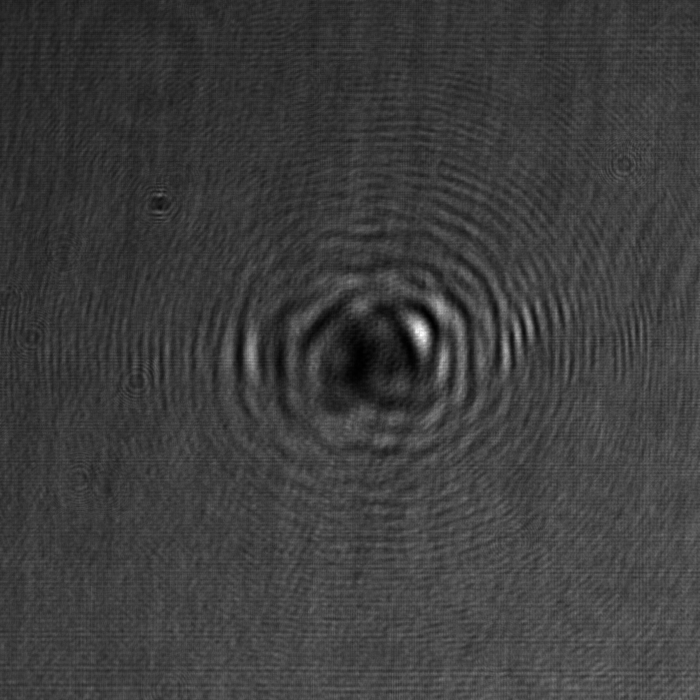}
		\caption{$t=12\Delta t$}
	\end{subfigure}\hspace{\gap}
	\begin{subfigure}[t]{\snapW}
		\centering
		\includegraphics[height=\snapH]{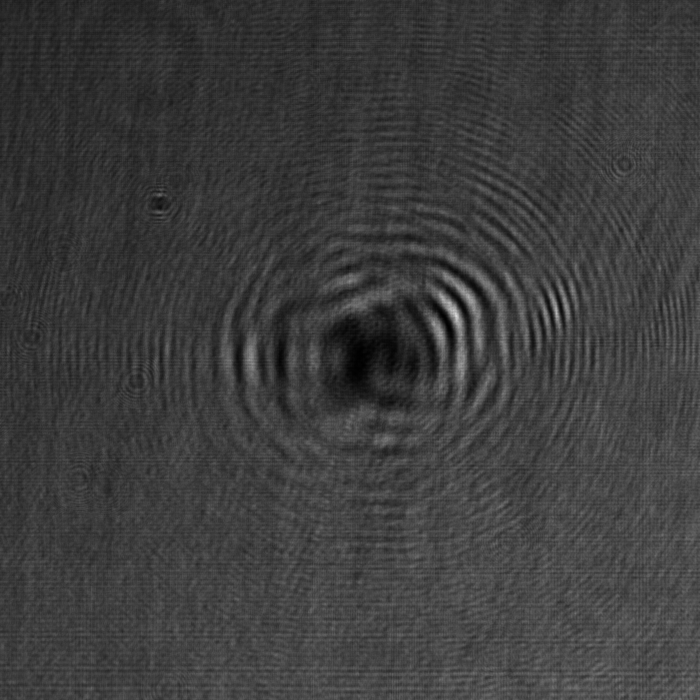}
		\caption{$t=18\Delta t$}
	\end{subfigure}\hspace{\gap}
	\begin{subfigure}[t]{\cbW}
		\centering
		\captionsetup{labelformat=empty} \includegraphics[height=\snapH]{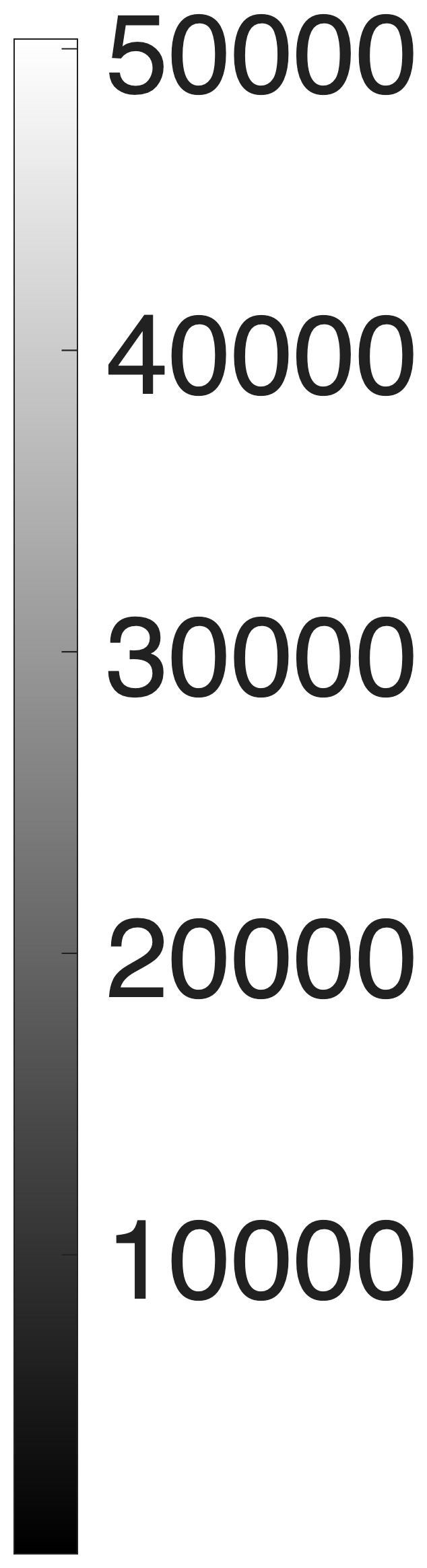}
\end{subfigure}
	
\begin{subfigure}[t]{\snapW}
		\centering
		\includegraphics[height=\snapH]{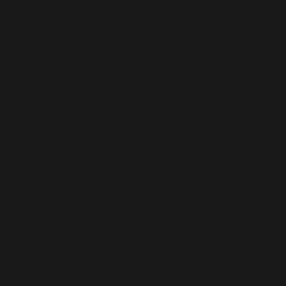}
		\caption{$t=0$}
	\end{subfigure}\hspace{\gap}
	\begin{subfigure}[t]{\snapW}
		\centering
		\includegraphics[height=\snapH]{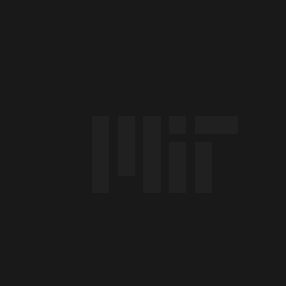}
		\caption{$t=1\Delta t$}
	\end{subfigure}\hspace{\gap}
	\begin{subfigure}[t]{\snapW}
		\centering
		\includegraphics[height=\snapH]{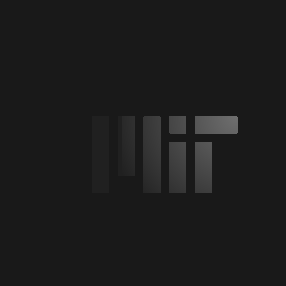}
		\caption{$t=7\Delta t$}
	\end{subfigure}\hspace{\gap}
	\begin{subfigure}[t]{\snapW}
		\centering
		\includegraphics[height=\snapH]{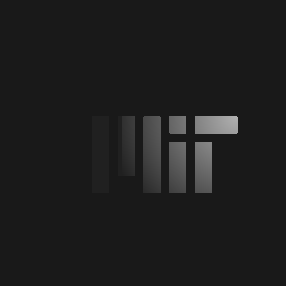}
		\caption{$t=12\Delta t$}
	\end{subfigure}\hspace{\gap}
	\begin{subfigure}[t]{\snapW}
		\centering
		\includegraphics[height=\snapH]{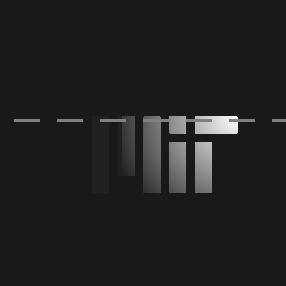}
		\caption{$t=18\Delta t$}
	\end{subfigure}\hspace{\gap}
	\begin{subfigure}[t]{\cbW}
		\centering
		\captionsetup{labelformat=empty} \includegraphics[height=\snapH]{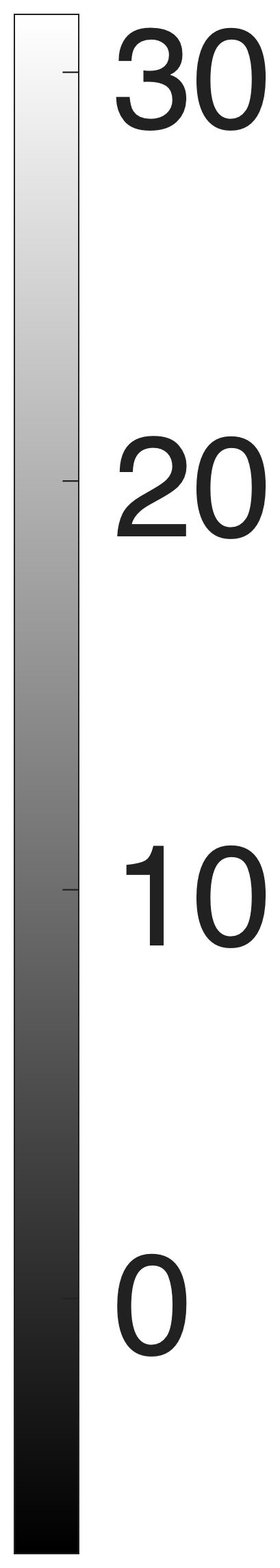}
\end{subfigure}

\begin{subfigure}[t]{\snapW}
		\centering
		\includegraphics[height=\snapH]{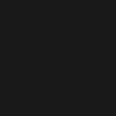}
		\caption{$t=0$}
		\label{fig:EXP-GrowingMIT/psi_xy0}
	\end{subfigure}\hspace{\gap}
	\begin{subfigure}[t]{\snapW}
		\centering
		\includegraphics[height=\snapH]{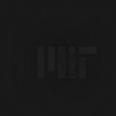}
		\caption{$t=1\Delta t$}
		\label{fig:EXP-GrowingMIT/psi_xyt_t03}
	\end{subfigure}\hspace{\gap}
	\begin{subfigure}[t]{\snapW}
		\centering
		\includegraphics[height=\snapH]{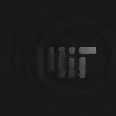}
		\caption{$t=7\Delta t$}
	\end{subfigure}\hspace{\gap}
	\begin{subfigure}[t]{\snapW}
		\centering
		\includegraphics[height=\snapH]{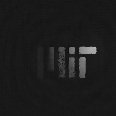}
		\caption{$t=12\Delta t$}
	\end{subfigure}\hspace{\gap}
	\begin{subfigure}[t]{\snapW}
		\centering
		\includegraphics[height=\snapH]{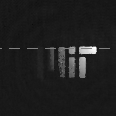}
		\caption{$t=18\Delta t$}
		\label{fig:EXP-GrowingMIT/psi_xyt_t18}
	\end{subfigure}\hspace{\gap}
	\begin{subfigure}[t]{\cbW}
		\centering
		\captionsetup{labelformat=empty} \includegraphics[height=\snapH]{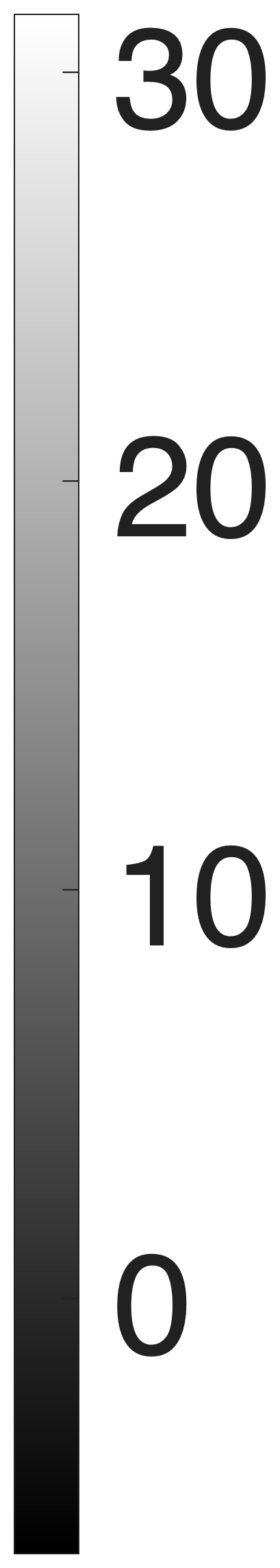}
\end{subfigure}

	\begin{subfigure}[t]{0.55\textwidth}
	\centering
	\includegraphics[width=\linewidth]{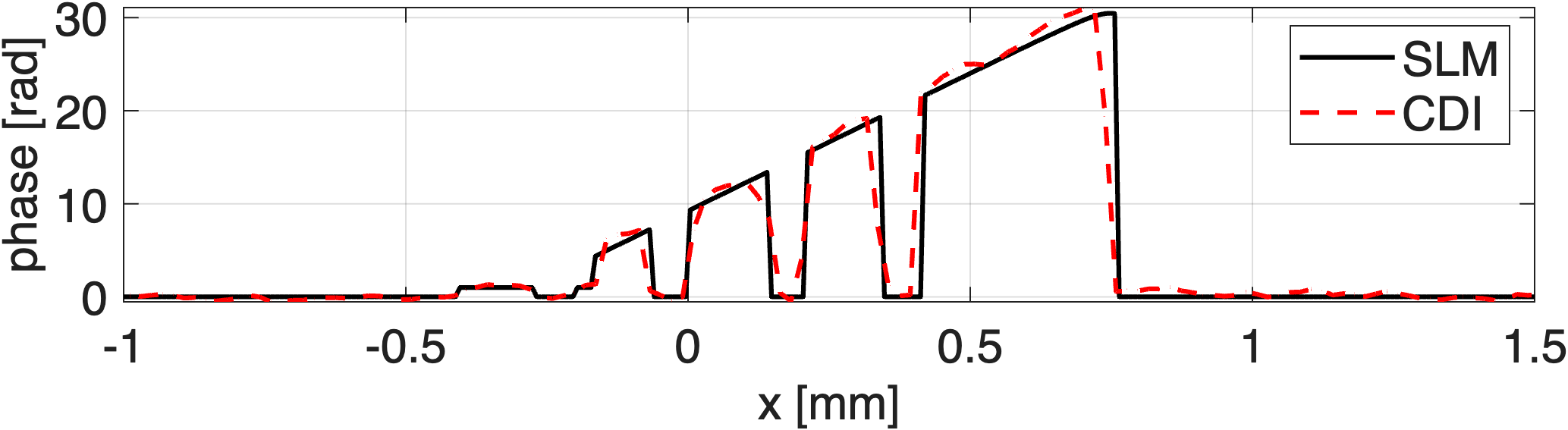}
	\caption{$t=18\Delta t$}
	\end{subfigure}

	\caption{
	Dynamic CDI reconstruction from an SLM-driven phase trajectory of pure growth, using $\alpha = 3\pi / 2$.
	Row~1: measured diffraction frames $I_{\mathrm{meas}}(x',y',t)$.
	Row~2: ground-truth object-plane phase $\phi_{\mathrm{SLM}}(x,y,t)$ displayed on the SLM at selected times, shown over the central region of the field of view, cropped by a factor of~3 relative to Row~1.
	Row~3: reconstructed object-plane phase $\phi(x,y,t)$ at the same times, obtained by integrating the reconstructed inter-frame updates from the estimated initial condition, displayed over the same cropped central region as Row~2.
	Panel~(p): cross-section comparison between the ground-truth phase and the CDI reconstruction at $t = 18\Delta t$. The cross-section was taken along the dashed lines shown in panels~(j) and~(o).
	Movies corresponding to rows 1--3 are available online. 
	}
	\label{fig:growing_MIT_result}
\end{figure*} 
Because $\Delta\psi(x,y,t)$ encodes only the frame-to-frame change, it is typically smaller in magnitude and more localized than the full field $\psi(x,y,t)$, so the temporal-increment constraint binds more tightly and the recovered $\Delta\phi(x,y,t)$ is less susceptible to noise. The integrated phase $\phi(x,y,t)$, by contrast, accumulates inter-frame errors over time, which may account for its noisier appearance in Fig.~\ref{fig:ReactionDiffusion_result}(l--o) relative to the inter-frame increments in Fig.~\ref{fig:ReactionDiffusion_result}(t--w). In settings where the dynamics itself is the quantity of interest, the inter-frame increment can be more informative than the accumulated phase.

The advantage of recovering the phase trajectory by temporal integration without explicit phase unwrapping (Sec.~\ref{sec:method-bound}) is shared by all three experiments, but is most striking when the accumulated phase exceeds $2\pi$. This is illustrated in the next experiment.

\subsection{Pure-growth dynamics}

In the first experiment, the object phase trajectory expands spatially over time. In the second, the spatial extent of the object remains fixed while the phase grows to a magnitude well beyond $2\pi$. A resized MIT logo on a fixed support was loaded onto the SLM as the ground-truth dynamics: starting from a uniformly zero-phase frame, the phase on the logo support grows monotonically in time, reaching a maximum value of $10\pi$ in the final frame. The phase increase is spatially nonuniform across the logo, and zero outside the logo support throughout the sequence. Implementation details of the prescribed growth profile are provided in Supplementary Material.

The reconstruction results are shown in Fig.~\ref{fig:growing_MIT_result}. The measured diffraction frames in Row~1 change substantially over time as the phase on the fixed support grows well beyond $2\pi$. Rows~2 and~3 show that the reconstructed object-plane phase tracks the ground truth at the selected times: the spatial pattern of the logo is recovered, and the phase values across the logo are consistent with the ground truth phase trajectory.

The cross-section in Fig.~\ref{fig:growing_MIT_result}(p) confirms the no-unwrapping property of the framework (Sec.~\ref{sec:method-bound}) at $t = 18\Delta t$: the discontinuities in the phase profile are recovered at the correct spatial locations, and the reconstructed phase values agree with the ground truth up to values near $10\pi$. This experiment also exhibits the error-accumulation limitation: small inter-frame errors add up over the forward and backward passes, producing concentric ring artifacts in Figs.~\ref{fig:growing_MIT_result}(m--o) that are absent from the corresponding ground-truth frames in Figs.~\ref{fig:growing_MIT_result}(h--j).

\subsection{In-situ metrology of photo-polymer reactions}

The third set of experiments applies the framework to in-situ monitoring of a photo-polymer 3D printing process. Unlike the SLM experiments, the phase trajectory is not known a priori: the sample is a liquid monomer that undergoes irreversible photo-chemical reaction under a spatially patterned blue printing beam, and the goal is to recover the spatiotemporal evolution of the phase profile induced by the printed sample as polymerization proceeds.

\begin{figure}[!htbp]
	\centering
	\includegraphics[width=0.45\textwidth]{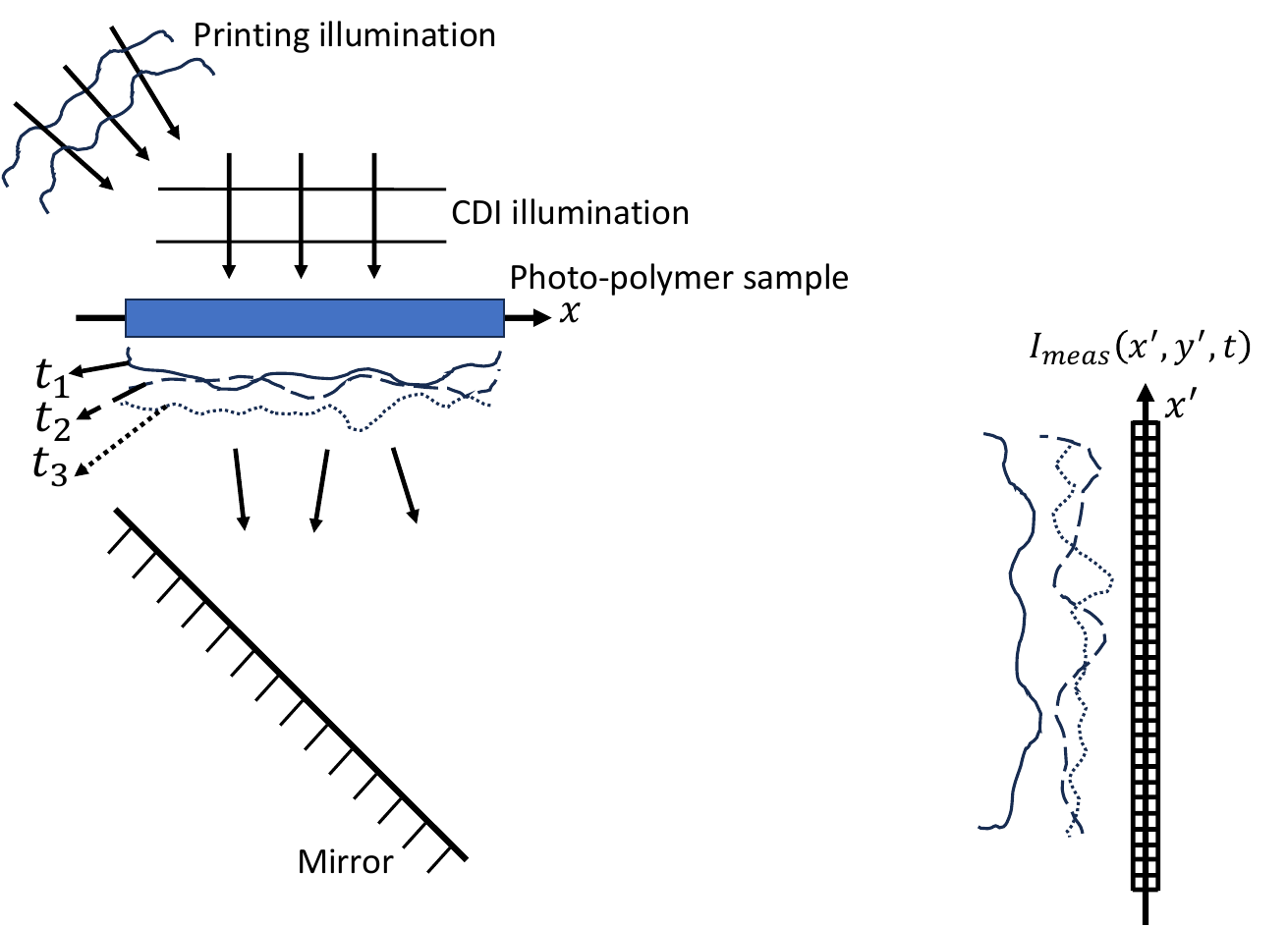}
	\caption{Measurement geometry for in-situ dynamic CDI of a photo-polymer
		3D printing process. The blue printing beam drives the photo-chemical
		reaction; the probe beam, at a spectrally separated wavelength,
		observes the resulting phase evolution via Fresnel diffraction recorded
		on the detector at distance $z$.}
	\label{fig:printing-measurement-geometry}
\end{figure} 
The measurement geometry is shown in Fig.~\ref{fig:printing-measurement-geometry}. Two beams are incident on the photo-polymer sample: a spatially patterned blue printing beam that drives the photo-chemical reaction, and a He--Ne probe beam whose Fresnel diffraction is recorded on the detector. The probe wavelength is spectrally separated from the printing wavelength, so the diffraction movie observes the evolving phase passively while the printing beam controls the dynamics. We present two printing experiments that differ in how the printing illumination is patterned in space and time. In the first, the printing beam is shaped by a circular aperture projection, and the aperture diameter is stepped through three discrete values during printing so that the illuminated region expands at prescribed times --- the printing analog of the reaction--diffusion experiment. In the second, the printing beam is shaped by a portion of a USAF resolution target held fixed in space throughout the print, so that only the phase magnitude on a fixed support evolves --- the printing analog of the pure-growth experiment.

\begin{figure*}[t]
	\centering
	
	\newcommand{\snapH}{2.35cm}      \newcommand{\snapW}{0.13\textwidth} \newcommand{\cbW}{0.01\textwidth}   \newcommand{\gap}{0.5pt}          

\begin{subfigure}[t]{\snapW}
		\centering
		\includegraphics[height=\snapH]{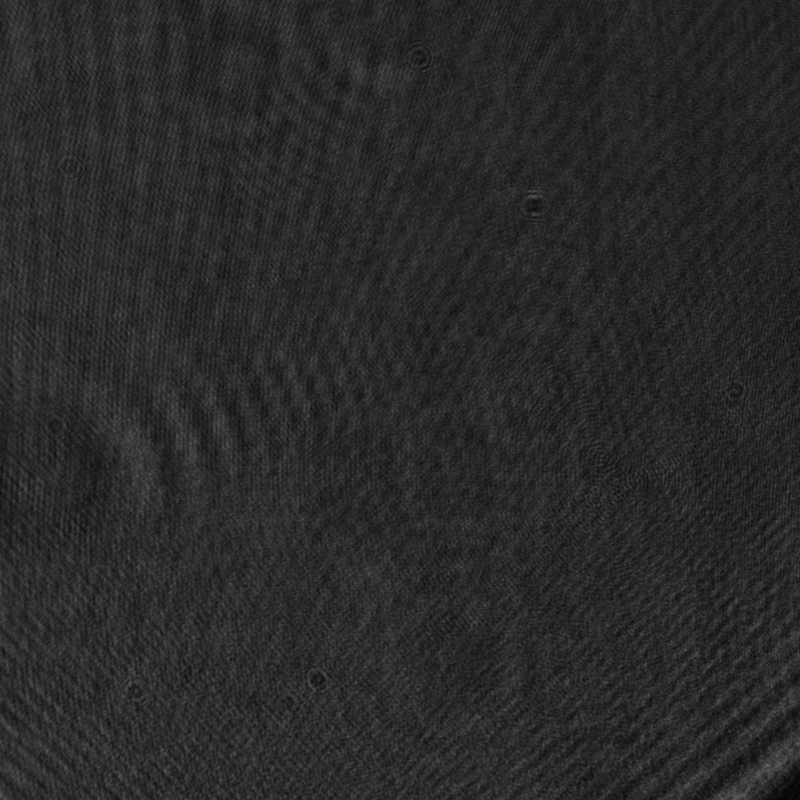}
		\caption{$t=0$}
		\label{fig:EXP-3DPrint/I_xy0}
	\end{subfigure}\hspace{\gap}
	\begin{subfigure}[t]{\snapW}
		\centering
		\includegraphics[height=\snapH]{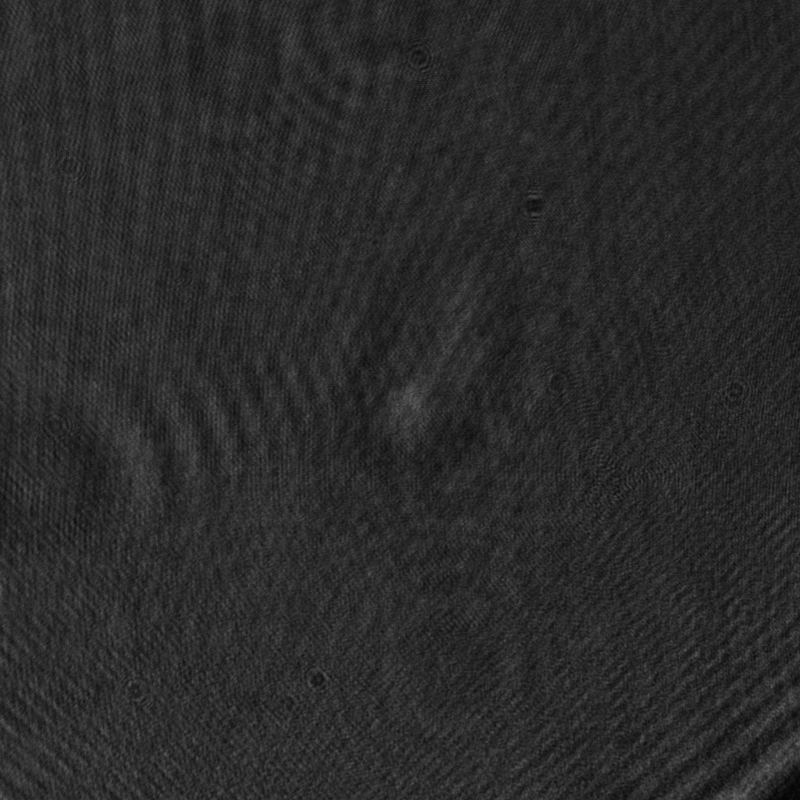}
		\caption{$t=2\Delta t$}
	\end{subfigure}\hspace{\gap}
	\begin{subfigure}[t]{\snapW}
		\centering
		\includegraphics[height=\snapH]{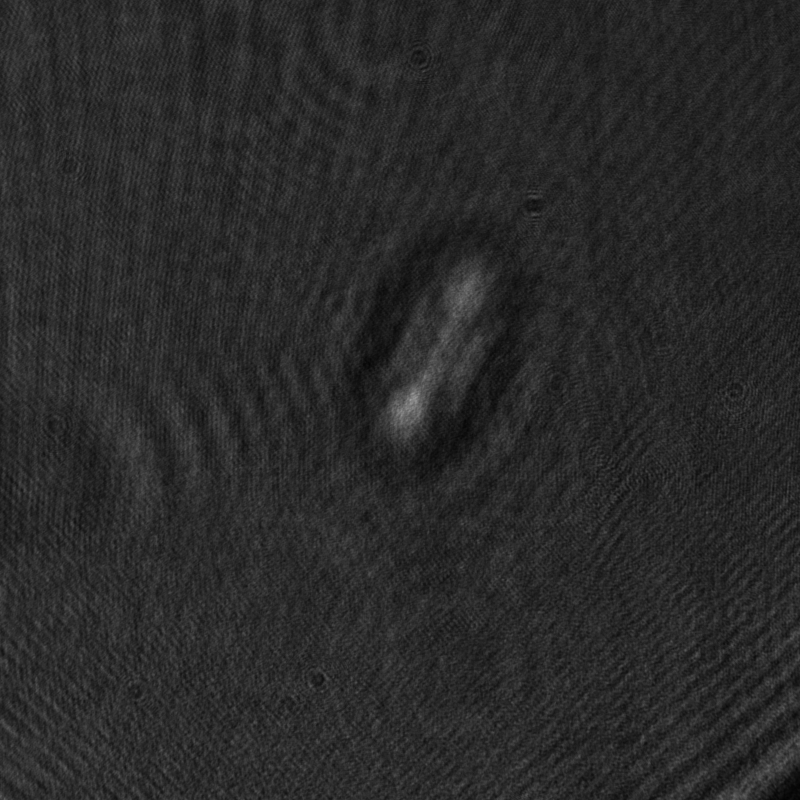}
		\caption{$t=4\Delta t$}
	\end{subfigure}\hspace{\gap}
	\begin{subfigure}[t]{\snapW}
		\centering
		\includegraphics[height=\snapH]{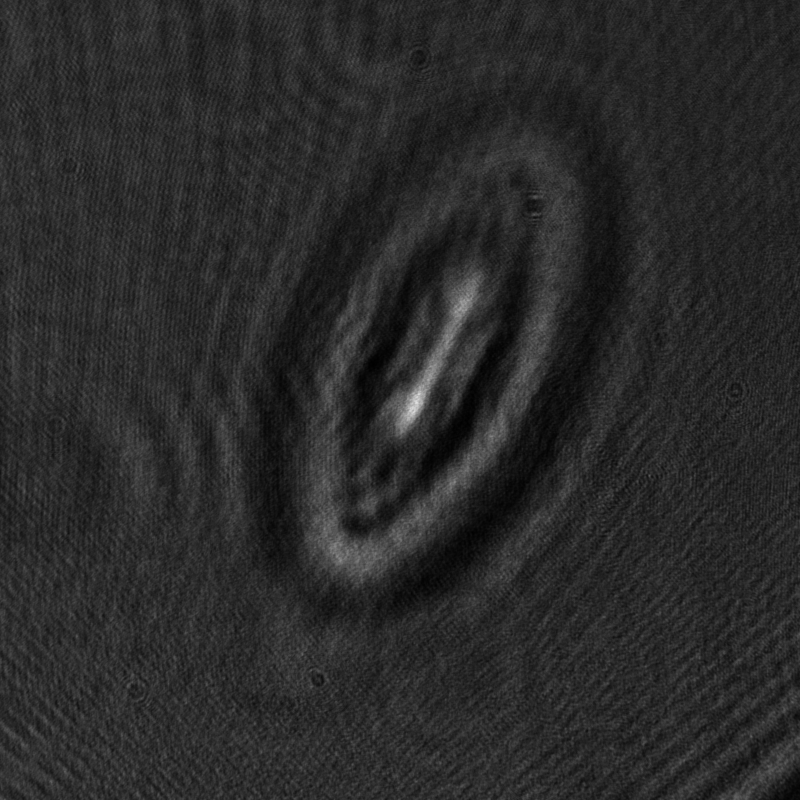}
		\caption{$t=8\Delta t$}
	\end{subfigure}\hspace{\gap}
	\begin{subfigure}[t]{\snapW}
		\centering
		\includegraphics[height=\snapH]{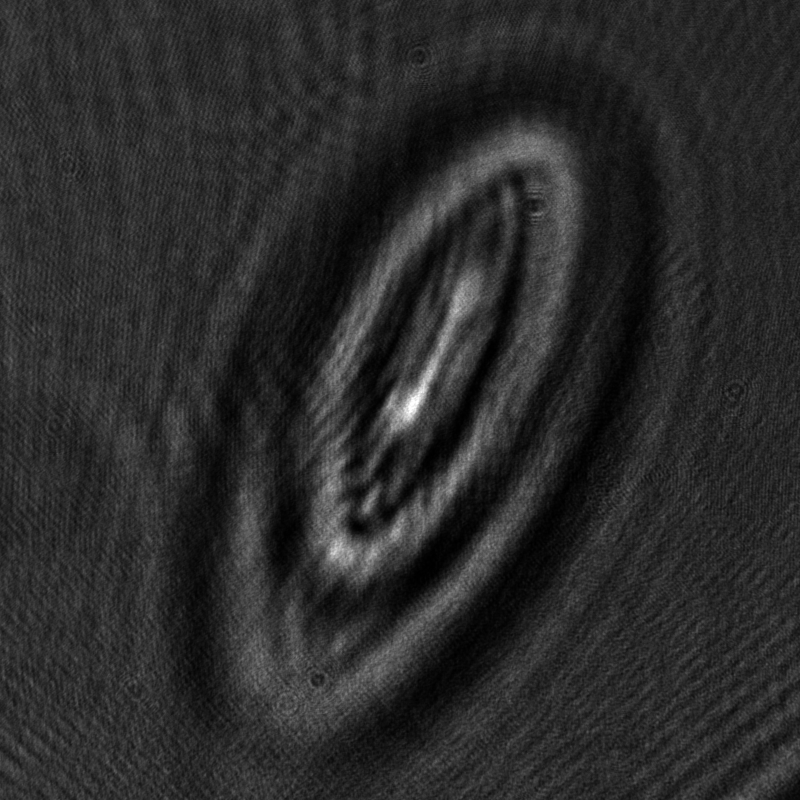}
		\caption{$t=15\Delta t$}
	\end{subfigure}\hspace{\gap}
	\begin{subfigure}[t]{\cbW}
		\centering
		\captionsetup{labelformat=empty} \includegraphics[height=\snapH]{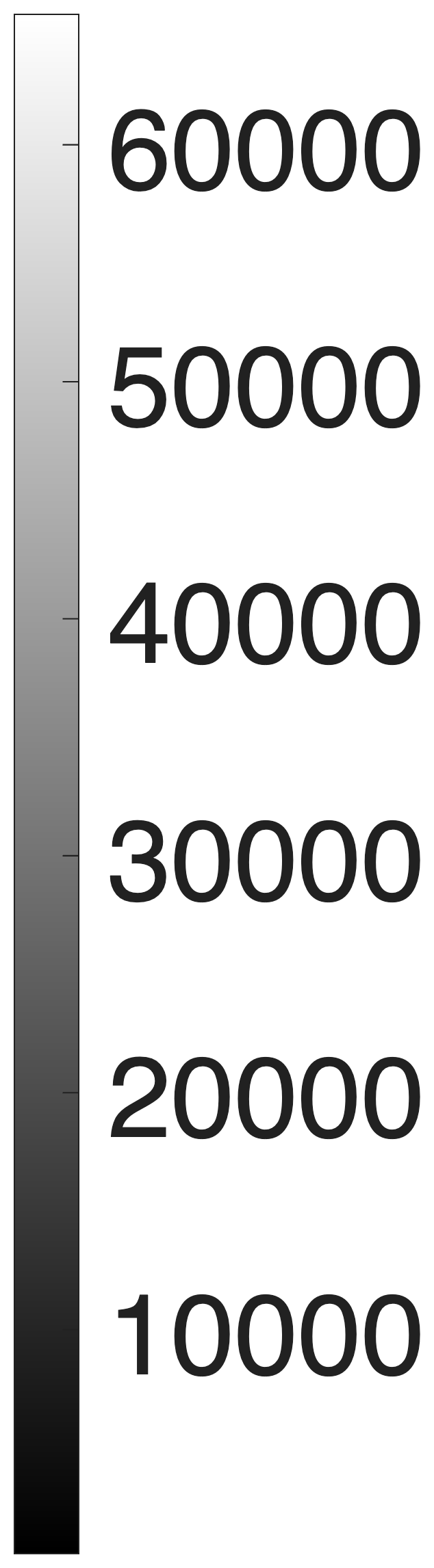}
\end{subfigure}

\begin{subfigure}[t]{\snapW}
		\centering
		\includegraphics[height=\snapH]{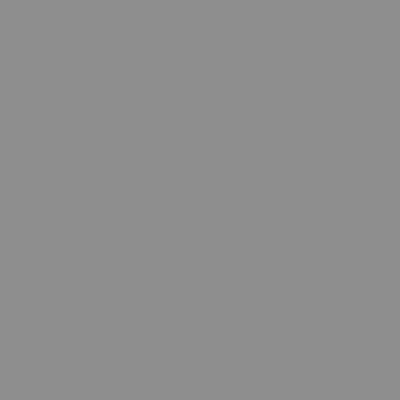}
		\caption{$t=0$}
		\label{fig:EXP-3DPrint/psi_xy0}
	\end{subfigure}\hspace{\gap}
	\begin{subfigure}[t]{\snapW}
		\centering
		\includegraphics[height=\snapH]{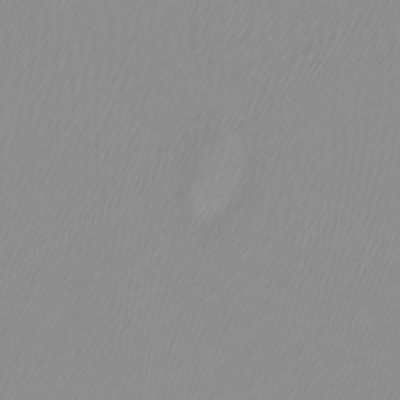}
		\caption{$t=1\Delta t$}
		\label{fig:EXP-3DPrint/psi_xyt_t03}
	\end{subfigure}\hspace{\gap}
	\begin{subfigure}[t]{\snapW}
		\centering
		\includegraphics[height=\snapH]{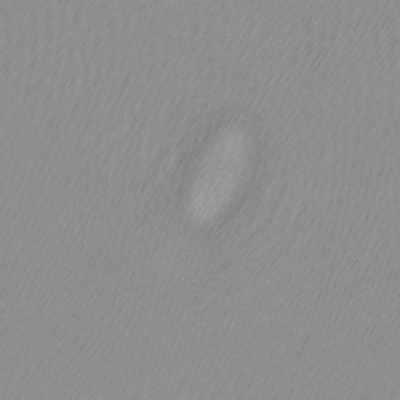}
		\caption{$t=7\Delta t$}
	\end{subfigure}\hspace{\gap}
	\begin{subfigure}[t]{\snapW}
		\centering
		\includegraphics[height=\snapH]{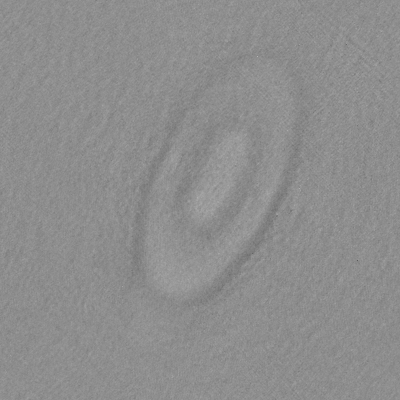}
		\caption{$t=12\Delta t$}
	\end{subfigure}\hspace{\gap}
	\begin{subfigure}[t]{\snapW}
		\centering
		\includegraphics[height=\snapH]{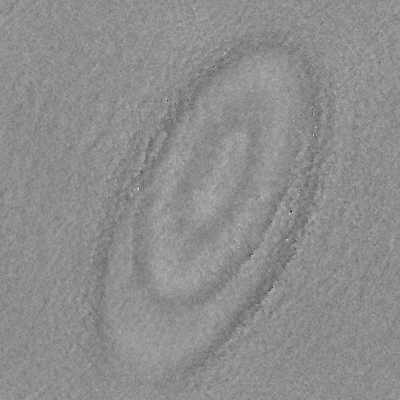}
		\caption{$t=18\Delta t$}
		\label{fig:EXP-3DPrint/psi_xyt_t18}
	\end{subfigure}\hspace{\gap}
	\begin{subfigure}[t]{\cbW}
		\centering
		\captionsetup{labelformat=empty} \includegraphics[height=\snapH]{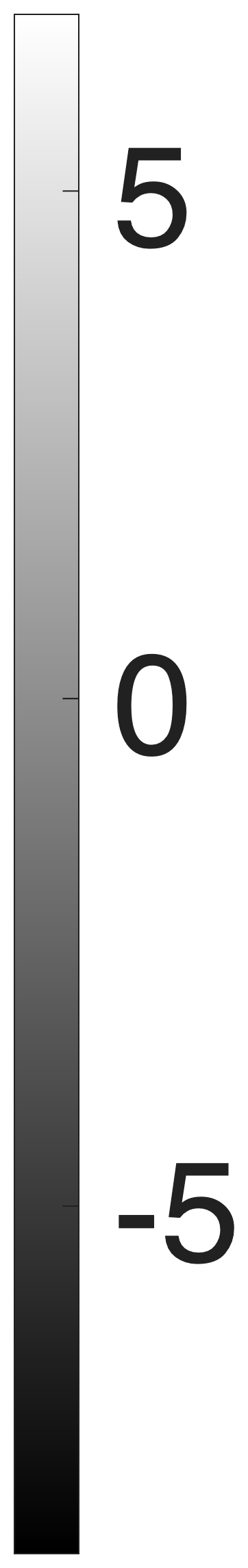}
\end{subfigure}

\caption{
	In-situ dynamic CDI reconstruction during photo-polymer 3D printing with a circular aperture stepped through three discrete diameters during fabrication.
	Row~1: measured diffraction frames $I_{\mathrm{meas}}(x',y',t)$.
	Row~2: reconstructed object-plane phase $\phi(x,y,t)$ at select times, obtained by integrating the inter-frame factors from the estimated initial condition.
	Movies corresponding to rows 1--2 are available online.
}
	\label{fig:3DPrint_result_expanding}
\end{figure*} 
Results from the first printing experiment are shown in Fig.~\ref{fig:3DPrint_result_expanding}. At $t = 0$, before the printing beam is turned on, the diffraction pattern in Fig.~\ref{fig:3DPrint_result_expanding}(a) corresponds to the unpolymerized monomer layer and the probe alone. As printing proceeds, the diffraction pattern develops increasingly prominent ring-like fringes that grow in spatial extent and contrast, with the three discrete jumps in aperture diameter visible as abrupt changes in the spatial scale of the diffraction features (Figs.~\ref{fig:3DPrint_result_expanding}(b--e)). The corresponding reconstructed sample-plane phase is shown in Figs.~\ref{fig:3DPrint_result_expanding}(f--j). The phase is spatially uniform at $t = 0$, and at subsequent times an elliptical phase structure emerges, consistent with the circular aperture projected at oblique incidence, and grows in magnitude monotonically within the printed region. The spatial extent of the polymerized region expands in step with the prescribed aperture diameter changes.

\begin{figure*}[!htbp]
	\centering
	
	\newcommand{\snapH}{2.35cm}      \newcommand{\snapW}{0.13\textwidth} \newcommand{\cbW}{0.01\textwidth}   \newcommand{\gap}{0.5pt}          

\begin{subfigure}[t]{\snapW}
		\centering
		\includegraphics[height=\snapH]{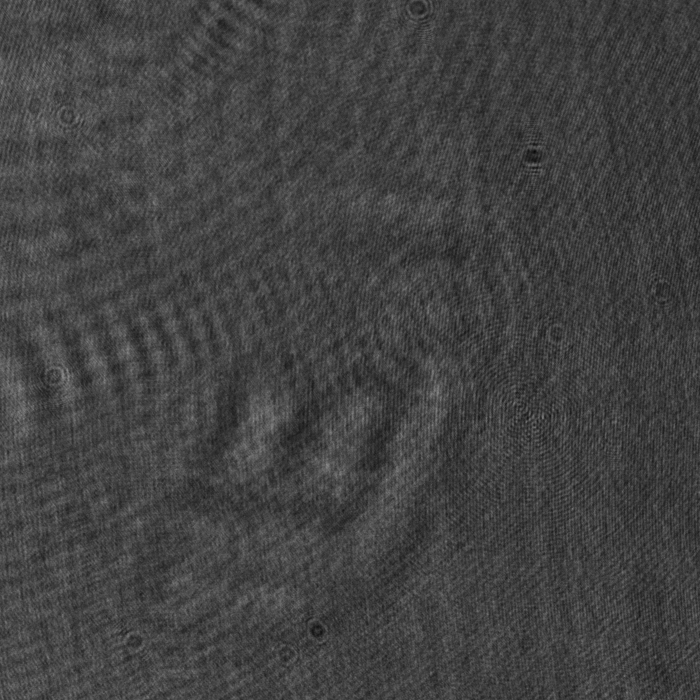}
		\caption{$t=0$}
		\label{fig:EXP-3DPrint/I_xy0}
	\end{subfigure}\hspace{\gap}
	\begin{subfigure}[t]{\snapW}
		\centering
		\includegraphics[height=\snapH]{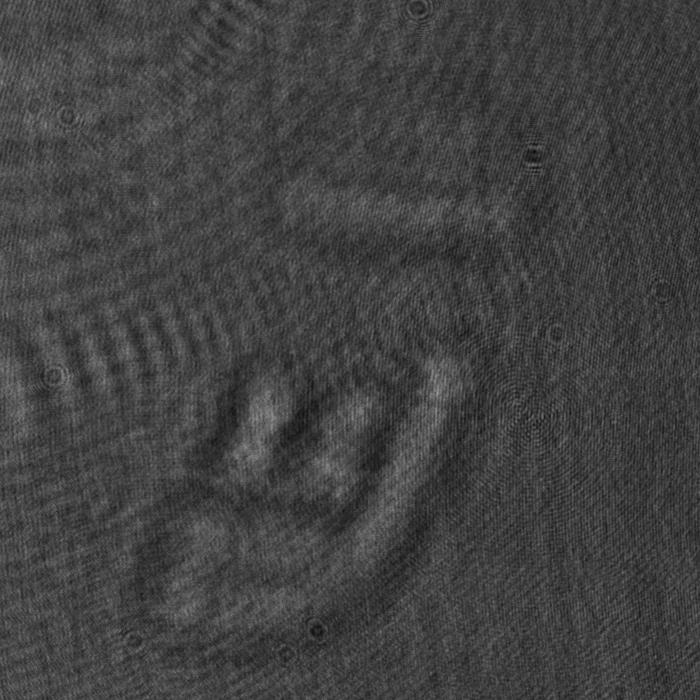}
		\caption{$t=2\Delta t$}
	\end{subfigure}\hspace{\gap}
	\begin{subfigure}[t]{\snapW}
		\centering
		\includegraphics[height=\snapH]{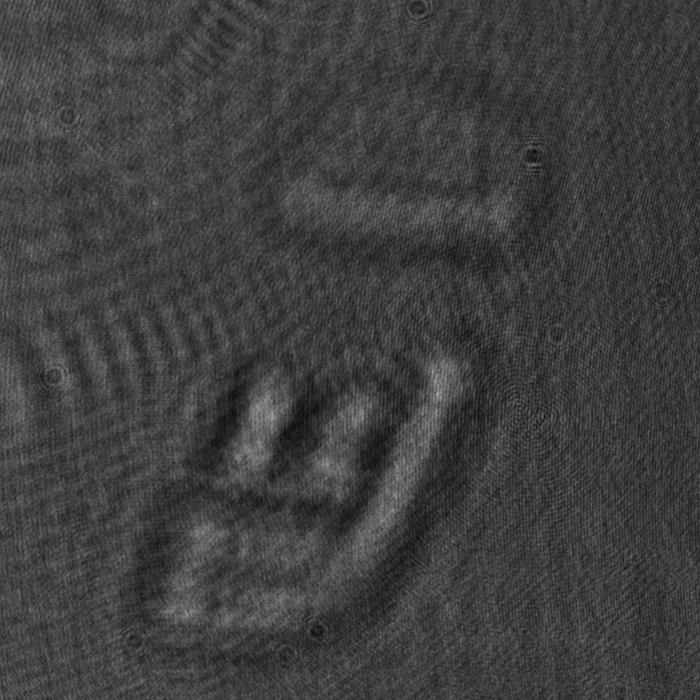}
		\caption{$t=4\Delta t$}
	\end{subfigure}\hspace{\gap}
	\begin{subfigure}[t]{\snapW}
		\centering
		\includegraphics[height=\snapH]{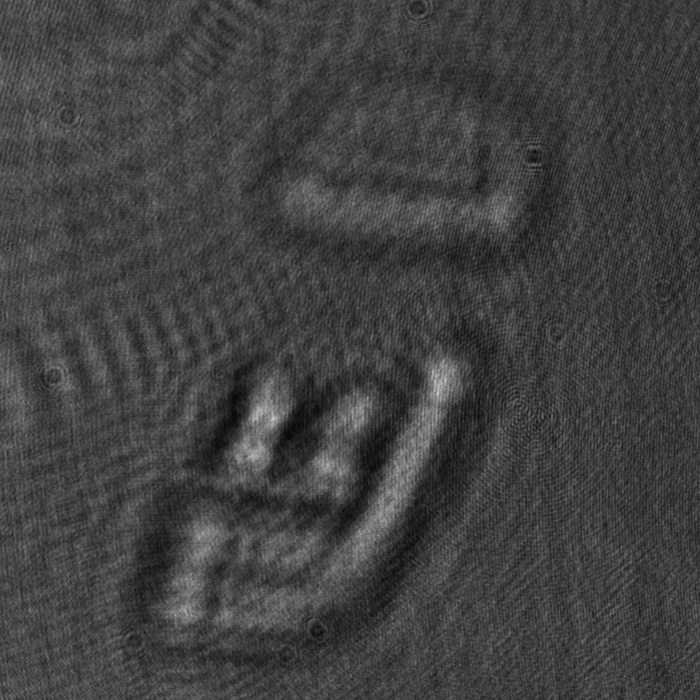}
		\caption{$t=8\Delta t$}
	\end{subfigure}\hspace{\gap}
	\begin{subfigure}[t]{\snapW}
		\centering
		\includegraphics[height=\snapH]{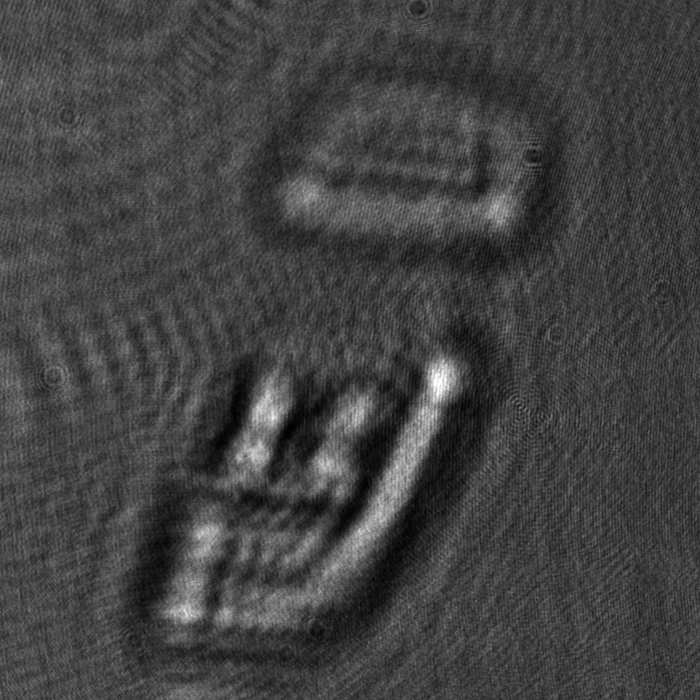}
		\caption{$t=15\Delta t$}
	\end{subfigure}\hspace{\gap}
	\begin{subfigure}[t]{\cbW}
		\centering
		\captionsetup{labelformat=empty} \includegraphics[height=\snapH]{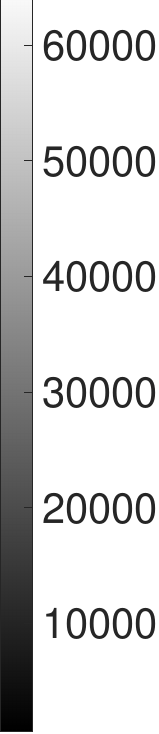}
\end{subfigure}

\begin{subfigure}[t]{\snapW}
		\centering
		\includegraphics[height=\snapH]{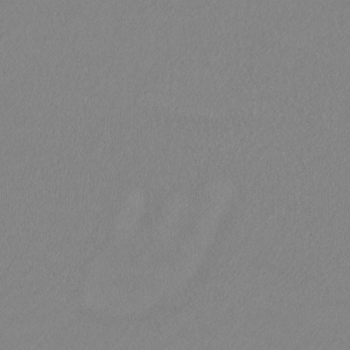}
		\caption{$t=0$}
		\label{fig:EXP-3DPrint/psi_xy0}
	\end{subfigure}\hspace{\gap}
	\begin{subfigure}[t]{\snapW}
		\centering
		\includegraphics[height=\snapH]{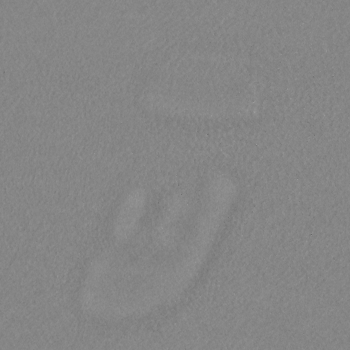}
		\caption{$t=1\Delta t$}
		\label{fig:EXP-3DPrint/psi_xyt_t03}
	\end{subfigure}\hspace{\gap}
	\begin{subfigure}[t]{\snapW}
		\centering
		\includegraphics[height=\snapH]{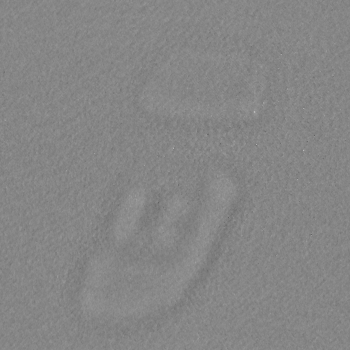}
		\caption{$t=7\Delta t$}
	\end{subfigure}\hspace{\gap}
	\begin{subfigure}[t]{\snapW}
		\centering
		\includegraphics[height=\snapH]{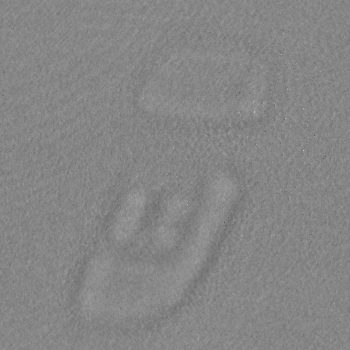}
		\caption{$t=12\Delta t$}
	\end{subfigure}\hspace{\gap}
	\begin{subfigure}[t]{\snapW}
		\centering
		\includegraphics[height=\snapH]{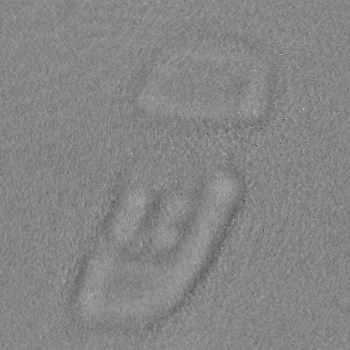}
		\caption{$t=18\Delta t$}
		\label{fig:EXP-3DPrint/psi_xyt_t18}
	\end{subfigure}\hspace{\gap}
	\begin{subfigure}[t]{\cbW}
		\centering
		\captionsetup{labelformat=empty} \includegraphics[height=\snapH]{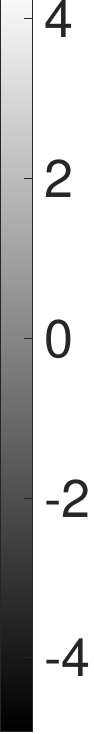}
\end{subfigure}

\caption{
	In-situ dynamic CDI reconstruction during photo-polymer 3D printing with a fixed USAF resolution target as the printing mask.
	Row~1: measured diffraction frames $I_{\mathrm{meas}}(x',y',t)$.
	Row~2: reconstructed object-plane phase $\phi(x,y,t)$ at select times, obtained by integrating the inter-frame factors from the estimated initial condition.
	Movies corresponding to rows 1--2 are available online.
}
	\label{fig:3DPrint_result_growing}
\end{figure*} 
Results from the second printing experiment are shown in Fig.~\ref{fig:3DPrint_result_growing}. The illuminated region is fixed throughout the print, so the only evolution is the monotonic growth of the phase magnitude on a fixed support. The reconstructed phase in Figs.~\ref{fig:3DPrint_result_growing}(f--j) shows the projected target pattern emerging and growing in magnitude over time, consistent with the progressive conversion of monomer to polymer within the illuminated region.

In both printing experiments, the reconstruction uses the same circular-sector constraint and the same algorithmic framework as the SLM experiments, with no additional priors specific to the printing process. The recovered phase trajectory provides a direct, quantitative observable of the spatiotemporal refractive-index evolution during printing. 	\section{Discussion}

We have shown that a temporal-increment-bound prior on the sample dynamics,
implemented as a bounded support constraint on the inter-frame factor in the
complex plane, is sufficient to reconstruct a time-varying phase trajectory
from a diffraction movie.
The constraint confines each pixel of the inter-frame factor to a circular
sector that simultaneously bounds the inter-frame phase increment and the
amplitude.
Together with the measurement-domain modulus constraint from the diffraction
movie, these two sets define a feasibility problem whose solution yields the
inter-frame factors from which the full phase trajectory is recovered by
temporal integration.

\vspace{0.5em}

Our implementation is deliberately minimalistic: the only priors are the modulus
constraint from the diffraction data and the circular-sector constraint, with
no additional spatial or temporal regularizers such as total variation,
optical flow, low-rank priors, or learned priors.
The fact that this minimalistic pair already yields quantitative reconstructions of phase trajectories
in three distinct experimental settings — including a real photo-chemical
process with no ground truth — suggests that a bounded support on the
inter-frame factor in the complex plane provides useful temporal
regularization on its own, without spatial priors of any kind.
A practical consequence is that the phase trajectory can be reconstructed
even when the total accumulated phase spans many $2\pi$ wraps, because
integration accumulates through a sequence of small, constrained increments
that each lie within $(-\pi,\pi)$ by construction, so phase unwrapping is
never required.

\vspace{0.5em}

The circular-sector constraint generalizes the classical nonnegativity
constraint \cite{Fienup1978ReconstructionModulusFT, Fienup1982PRComparison,
	TianFienup2023NonnegativityOnlyPR} and is related to histogram constraints in
crystallography \cite{Elser2003PhaseRetrievalIteratedProjections}; similar
complex-plane constraints have been used in X-ray imaging
\cite{Clark2010ComplexConstraintCDI, ChenFannjiang2018DRSingleMask}.
Nonnegativity is the degenerate case of a sector on the positive real axis,
while histogram constraints prescribe the full value distribution on the real
line.
The circular-sector constraint instead prescribes only the support of the
pixel-value distribution in the complex plane, with no dependence on pixel
location, local gradients, or spatially stationary regions.
This distinguishes it from recent dynamic CDI approaches that impose
spatially structured temporal constraints, such as adaptive stationary-region
priors.
The spatial information needed to disambiguate solutions is provided entirely
by the diffraction forward model, which couples all object-plane pixels
through the Fresnel integral.
We hypothesize that this separation — spatial coupling through the forward
model, temporal coupling through the inter-frame parameterization — is what
makes the approach effective with so few priors.
Combining the circular-sector constraint with spatially structured priors
may improve robustness in regimes where the present method struggles, but is
beyond the scope of this paper.

\vspace{0.5em}

The 3D printing experiments suggest a direct application of the framework
beyond imaging.
The recovered phase trajectory is a quantitative map of the spatiotemporal
refractive-index evolution during printing, and constitutes the observable
needed for photo-chemical system identification: fitting the kinetic
parameters of the reaction--diffusion PDE governing polymerization to the
measured phase history.
This inverse problem is distinct from the phase retrieval problem solved
here, but the two can be mutually beneficial: dynamic phase retrieval as
presented here provides the data for system identification, and system
identification provides a model that can further regularize the phase
retrieval solution.
A joint formulation is a direction for future work.
The same logic applies to other fabrication or growth processes that produce
a temporally smooth phase object under coherent illumination, including
thin-film deposition, etching, and crystal growth, provided the inter-frame
phase increment remains within the sector bound set by the frame rate.

\vspace{0.5em}

Several refinements of the framework are possible.
The circular sector itself can be tightened adaptively, in the spirit of
ShrinkWrap \cite{MarchesiniHe2003ShrinkWrap,
	Chapman2006HighResolution3DXRAY}: starting from a loose sector, one updates
the support of the pixel-value distribution from the current iterate and
tightens the bound progressively until convergence degrades, providing an
automatic estimate of the inter-frame increment range without prior knowledge.
A more aggressive refinement would let the support take an arbitrary shape
adapted to the empirical pixel-value distribution; this is generally
non-convex, and the corresponding projection is neither unique nor guaranteed
to decrease the feasibility gap, so the iteration may become unstable.
A separate direction is to revisit the algorithmic choice itself: the
projection-based feasibility formulation used here is one of several options,
and a systematic comparison with nonlinear optimization formulations remains
to be carried out.
The inter-frame parameterization is also compatible with measurement
modalities that directly record temporal differences, such as event-based
sensors; defining the forward model directly in terms of the inter-frame
factor in that setting is a natural extension.

\vspace{0.5em}

One limitation of the approach is error accumulation through
temporal integration: small residual errors in the inter-frame estimates add
coherently over time, so the reconstructed trajectory drifts progressively
away from the truth at later frames, even when each increment is individually
well recovered.
This drift is visible in both SLM experiments at the latest frames.
Strategies that enforce global consistency across the full trajectory, rather
than estimating one increment at a time and integrating, may reduce this
drift and are a natural direction for future work. 	\begin{backmatter}
	
	\bmsection{Funding}
	Air Force Office of Scientific Research (FA9550-23-1-0284); MIT--Fujikura Partnership Fund; Ralph E.\ and Eloise F.\ Cross endowed chair in Manufacturing.
	
	\bmsection{Acknowledgment}
	Y.T., Y.C.K.C., H.D.E., and G.B.\ were supported by the Air Force Office of Scientific Research through the Multi-University Research Initiative (MURI) ``Searching for what's new: the systematic development of dynamic x-ray microscopy.'' T.T.\ and Y.K.\ acknowledge the MIT--Fujikura Partnership Fund. G.B.\ acknowledges the Ralph E.\ and Eloise F.\ Cross endowed chair in Manufacturing. Y.T.\ thanks Difei Zhang at MIT EECS and Wenle Yan at MIT DMSE for helpful discussions. The authors acknowledge Claude, an AI, for assistance; the authors have reviewed and assume full responsibility for the results of the AI's work.
	
	\bmsection{Disclosures}
	T.T., Y.K.: Fujikura Ltd. (E). The remaining authors declare no conflicts of interest.
	
	\bmsection{Data availability}
	Data and code underlying the results presented in this paper are available in Ref.~\cite{TianCDIcode2026}.
	
\end{backmatter} 	\appendix

\section{Projection onto the circular sector}\label{sec:appendix-Pio}

The object-domain projector $\Pi_o$ defined in \eqref{eq:Pi_o_def} admits the following closed form. Writing $w = \Delta\psi(x,y,t) = re^{i\theta}$,
\begin{equation}
	\begin{aligned}
		\rho_{+}(w)
		&=\min\{\max\{\Re(w e^{-i\alpha/2}),0\},r_{\max}\},\\
		\rho_{-}(w)
		&=\min\{\max\{\Re(w e^{+i\alpha/2}),0\},r_{\max}\},\\[4pt]
		\Pi_o(w)
		&=
		\begin{cases}
			w,
			& r\le r_{\max},\ |\theta|\le \alpha/2,\\[4pt]
			\dfrac{r_{\max}}{r}w,
			& r>r_{\max},\ |\theta|\le \alpha/2,\\[6pt]
			\rho_{+}(w)e^{i\alpha/2},
			& \theta>\alpha/2,\\[6pt]
			\rho_{-}(w)e^{-i\alpha/2},
			& \theta<-\alpha/2.
		\end{cases}
	\end{aligned}
	\label{eq:Pi_o_closed_form}
\end{equation}
The four cases have a direct geometric interpretation. A pixel already inside $\mathcal{C}_\alpha$ is left unchanged (case~1). A pixel within the correct phase range but with amplitude exceeding $r_{\text{max}}$ is projected radially onto the arc $r = r_{\text{max}}$: its phase is preserved and its amplitude is clamped (case~2). A pixel whose phase falls outside $[-\alpha/2, \alpha/2]$ is projected orthogonally onto the nearer bounding ray, $\theta = +\alpha/2$ or $\theta = -\alpha/2$, and then has its component along that ray clamped to $[0, r_{\text{max}}]$ via the auxiliary functions $\rho_{\pm}(w)$ (cases~3--4). Geometrically, the complex value is rotated toward the sector edge until it contacts the boundary, and $\rho_{\pm}(w)$ then enforces two degenerate sub-cases: a negative component maps the pixel to the origin, and a component exceeding $r_{\text{max}}$ maps it to the sector corner $r_{\text{max}}\,e^{\pm i\alpha/2}$.

Applying \eqref{eq:Pi_o_closed_form} to each pixel of $\Delta\psi(x,y,t)$ for each $t$ yields $\Delta\psi_o(x,y,t)$. The radius cap is $r_{\text{max}}=1$ for $\Delta\psi$, since the thin-object model forbids amplitude amplification at the object plane. For the probe $\psi(x,y,0)$, $r_{\text{max}}$ is instead determined by the illumination intensity and is treated as a free parameter, as described in Sec.~\ref{sec:algo}.
\section{Measurement-domain operator on inter-frame factors}\label{sec:appendix-Pim}

We first state the $\ell_2$ projection form of $\Pi_m$ on the field $\psi$, and then extend it to act on the inter-frame factor $\Delta\psi$.

\paragraph{$\ell_2$ projection onto magnitude set.}
Since the Fresnel propagator $\mathcal{P}$ defined in \eqref{eq:fresnel_integral} is unitary, conjugating the detector-plane magnitude swap \eqref{eq:mag_replace} with $\mathcal{P}$ and $\mathcal{P}^{-1}$ preserves the Euclidean distance.
The three-step procedure of \eqref{eq:fwd_prop}--\eqref{eq:bwd_prop} is therefore equally the Euclidean projection in the object plane,
\begin{equation}
	\psi_m(\cdot,\cdot,t)
	=
	\argmin_{\phi:\,|\mathcal{P}\{\phi\}|=\sqrt{I(\cdot,\cdot,t)}}
	\|\phi - \psi(\cdot,\cdot,t)\|_2^2.
	\label{eq:Pi_m_as_projection}
\end{equation}
The subscript $m$ indicates that $\psi_m(x,y,t)$ is consistent with the intensity measurement $I(x',y',t)$.

\paragraph{Forward-pass operator.}
The forward-pass operator $\Pi^{\Delta}_{m,\mathrm{fwd}}$ acts on the inter-frame factor $\Delta\psi(x,y,(j-1)\Delta t)$ with the previous field $\psi(x,y,(j-1)\Delta t)$ treated as fixed.
Given a current iterate $\Delta\psi$, the next frame is formed via \eqref{eq:psi_j_from_delta}.
Applying $\Pi_m$ to this product yields $\psi_m(x,y,j\Delta t)$, the minimal change to $\psi(x,y,j\Delta t)$ that enforces consistency with $I(x',y',j\Delta t)$.

One method to recover the corresponding inter-frame factor would be to divide $\psi_m(x,y,j\Delta t)$ pixelwise by the fixed previous field $\psi(x,y,(j-1)\Delta t)$.
However, this division becomes ill-conditioned where $\psi(x,y,(j-1)\Delta t)$ has small amplitude --- a situation that arises generically at the dark regions of the probe and at the dark regions of the evolving sample.
A Tikhonov-regularized division can stabilize the inversion, but only by introducing a pixelwise bias that depends on $|\psi(x,y,(j-1)\Delta t)|$ and therefore acts unevenly across the frame.

We instead update $\Delta\psi$ via gradient descent on the least-squares mismatch, in the spirit of the ePIE algorithm \cite{MaidenRodenburg2009PIE}, which avoids any pointwise division.
We seek the $\Delta\psi$ that makes the evolved next frame $\psi(x,y,(j-1)\Delta t)\,\Delta\psi(x,y,(j-1)\Delta t)$ match the measurement-consistent target $\psi_m(x,y,j\Delta t)$ in the least-squares sense:
\begin{equation}
	\mathcal{J}(\Delta\psi)
	=
	\tfrac{1}{2}\bigl\|\,
	\psi(x,y,(j-1)\Delta t)\,\Delta\psi
	\;-\;
	\psi_m(x,y,j\Delta t)
	\,\bigr\|_2^2.
	\label{eq:fwd_objective}
\end{equation}
With $\psi(x,y,(j-1)\Delta t)$ fixed, $\mathcal{J}$ is a quadratic in $\Delta\psi$, separable across pixels.
Its gradient with respect to $\Delta\psi^*$ is
\begin{equation}
	\frac{\partial\mathcal{J}}{\partial\Delta\psi^*}
	=
	\psi^*(x,y,(j-1)\Delta t)
	\bigl[\,
	\psi(x,y,(j-1)\Delta t)\,\Delta\psi
	\;-\;
	\psi_m(x,y,j\Delta t)
	\,\bigr].
	\label{eq:fwd_gradient}
\end{equation}
A gradient descent step with step size $\tau > 0$ reads
\begin{equation}
	\begin{aligned}
		\Delta\psi
		\;\leftarrow\;
		&\Delta\psi
		\;+\;
		\tau\,\psi^*(x,y,(j-1)\Delta t)
		\\
		&\times
		\bigl[\,
		\psi_m(x,y,j\Delta t)
		\;-\;
		\psi(x,y,(j-1)\Delta t)\,\Delta\psi
		\,\bigr].
	\end{aligned}
	\label{eq:fwd_gd_generic}
\end{equation}
Note that Hessian entry of \eqref{eq:fwd_objective}, $\partial^2\mathcal{J}/\partial\Delta\psi^*(x,y)\,\partial\Delta\psi(x',y')$, vanishes whenever $(x,y) \ne (x',y')$ because pixel $(x,y)$ of $\Delta\psi$ contributes only to the $(x,y)$ term of the sum.
The Hessian of $\mathcal{J}$ is thus diagonal in pixel space, with diagonal entries as
\begin{equation}
	\frac{\partial^2\mathcal{J}}{\partial\Delta\psi^*(x,y)\,\partial\Delta\psi(x,y)}
	=
	|\psi(x,y,(j-1)\Delta t)|^2.
\end{equation}
The largest monotone-stable step size is the reciprocal of the spectral radius of the Hessian, namely $\bigl(\max_{x,y}|\psi(x,y,(j-1)\Delta t)|^2\bigr)^{-1}$.
We therefore set
\begin{equation}
	\tau
	\;=\;
	\frac{\gamma}{\max_{x,y}|\psi(x,y,(j-1)\Delta t)|^2},
	\qquad
	\gamma \in (0,1],
	\label{eq:fwd_step}
\end{equation}
giving the update
\begin{equation}
	\begin{aligned}
		&\Delta\psi_m(x,y,(j-1)\Delta t)
		=
		\Delta\psi
		\\
		&+\;
		\gamma\,
		\frac{\psi^*(x,y,(j-1)\Delta t)\bigl[\,\psi_m(x,y,j\Delta t) - \psi(x,y,(j-1)\Delta t)\,\Delta\psi\,\bigr]}
		{\max_{x,y}|\psi(x,y,(j-1)\Delta t)|^2}.
	\end{aligned}
	\label{eq:delta_psi_m_forward}
\end{equation}
The denominator in \eqref{eq:delta_psi_m_forward} is a single scalar per frame rather than a pixelwise quantity, so the update applies a uniformly scaled correction across the whole frame.
No pointwise division by $\psi(x,y,(j-1)\Delta t)$ ever appears, so the dark regions of the previous field do not destabilize the recovery of $\Delta\psi$.
The map $\Delta\psi \mapsto \Delta\psi_m$ defined by \eqref{eq:psi_j_from_delta} and \eqref{eq:delta_psi_m_forward} is the forward-pass measurement operator $\Pi^{\Delta}_{m,\mathrm{fwd}}$.
In practice we apply several gradient steps of \eqref{eq:delta_psi_m_forward} with the target $\psi_m(x,y,j\Delta t)$ held fixed before passing $\Delta\psi_m$ to the object-domain projector, following the inner-iteration strategy of ePIE \cite{MaidenRodenburg2009PIE}.

\paragraph{Backward-pass operator.}
The backward-pass operator $\Pi^{\Delta}_{m,\mathrm{bwd}}$ acts on $\Delta\psi(x,y,(j-1)\Delta t)$ with the next field $\psi(x,y,j\Delta t)$ treated as fixed.
The construction parallels the forward-pass case but with the roles of the consecutive frames swapped.
We first form a current estimate of the previous field from the multiplicative coupling \eqref{eq:psi_j_from_delta},
\begin{equation}
	\widetilde{\psi}(x,y,(j-1)\Delta t)
	=
	\frac{\psi(x,y,j\Delta t)\,\Delta\psi^*(x,y,(j-1)\Delta t)
		\;+\;\epsilon^2}
	{|\Delta\psi(x,y,(j-1)\Delta t)|^2 \;+\;\epsilon^2},
	\label{eq:psi_prev_from_delta}
\end{equation}
where $\epsilon > 0$ is a small constant that biases $\widetilde{\psi}$ toward $\psi(x,y,j\Delta t)$ in pixels where $\Delta\psi$ is near zero.
Because $|\Delta\psi| \le 1$ everywhere by the passivity bound and is typically close to unity, this Tikhonov division is well-conditioned in practice.
Applying $\Pi_m$ to $\widetilde{\psi}(x,y,(j-1)\Delta t)$ --- this time enforcing consistency with the measurement $I(x',y',(j-1)\Delta t)$ --- gives the measurement-consistent estimate $\psi_m(x,y,(j-1)\Delta t)$.

We then update $\Delta\psi$ such that $\psi_m(x,y,(j-1)\Delta t)\,\Delta\psi$ matches the fixed next field $\psi(x,y,j\Delta t)$ in the least-squares sense:
\begin{equation}
	\mathcal{J}(\Delta\psi)
	=
	\tfrac{1}{2}\bigl\|\,
	\psi_m(x,y,(j-1)\Delta t)\,\Delta\psi
	\;-\;
	\psi(x,y,j\Delta t)
	\,\bigr\|_2^2.
	\label{eq:bwd_objective}
\end{equation}
The same gradient calculation that produced \eqref{eq:delta_psi_m_forward} now gives
\begin{equation}
	\begin{aligned}
		&\Delta\psi_m(x,y,(j-1)\Delta t)
		=
		\Delta\psi
		\\
		&+\;
		\gamma\,
		\frac{\psi_m^*(x,y,(j-1)\Delta t)\bigl[\,\psi(x,y,j\Delta t) - \psi_m(x,y,(j-1)\Delta t)\,\Delta\psi\,\bigr]}
		{\max_{x,y}|\psi_m(x,y,(j-1)\Delta t)|^2}.
	\end{aligned}
	\label{eq:delta_psi_m_backward}
\end{equation}
As in the forward-pass case, the denominator in \eqref{eq:delta_psi_m_backward} is a single scalar per frame, so the correction is applied uniformly across pixels.
The map $\Delta\psi \mapsto \Delta\psi_m$ defined by \eqref{eq:psi_prev_from_delta}--\eqref{eq:delta_psi_m_backward} is the backward-pass measurement operator $\Pi^{\Delta}_{m,\mathrm{bwd}}$.

\section{Reflectors and projection iteration maps}\label{sec:appendix-iter}

For projectors $\Pi_\mathcal{A}$ and $\Pi_\mathcal{B}$, the associated reflectors are
\begin{equation}
	\mathcal{R}_\mathcal{A} = 2\Pi_\mathcal{A} - I,
	\qquad
	\mathcal{R}_\mathcal{B} = 2\Pi_\mathcal{B} - I.
	\label{eq:reflectors}
\end{equation}
The reflector $\mathcal{R}_\mathcal{A}(x)$ maps $x$ to the point on the opposite side of $\mathcal{A}$ at the same distance, with $\Pi_\mathcal{A}(x)$ as the midpoint, so that $\Pi_\mathcal{A}(x) = \tfrac{1}{2}(x + \mathcal{R}_\mathcal{A}(x))$. Plain alternating projections can stagnate at fixed points outside $\mathcal{A} \cap \mathcal{B}$, particularly when the constraint sets are non-convex, as is the case for the measurement-magnitude constraint in Sec.~\ref{sec:method-projectors}. Using reflections at the intermediate step causes iterates to overshoot past the constraint boundary, which destabilizes such spurious fixed points.

The relaxed-reflect-reflect (RRR) map is
\begin{equation}
	\mathcal{T}_{\mathrm{RRR}}(x)
	=
	x + \beta\Bigl(
	\Pi_\mathcal{A}\!\bigl(\mathcal{R}_\mathcal{B}(x)\bigr)
	- \Pi_\mathcal{B}(x)
	\Bigr),
	\label{eq:RRR_map}
\end{equation}
and the relaxed averaged alternating reflections (RAAR) map is
\begin{equation}
	\mathcal{T}_{\mathrm{RAAR}}(x)
	=
	\frac{\beta}{2}\Bigl(
	\mathcal{R}_\mathcal{A}\mathcal{R}_\mathcal{B}(x) + x
	\Bigr)
	+ (1-\beta)\,\Pi_\mathcal{B}(x),
	\label{eq:RAAR_map}
\end{equation}
with relaxation parameter $\beta \in (0,1]$. The fixed points of $\mathcal{T}_{\mathrm{RRR}}$ and $\mathcal{T}_{\mathrm{RAAR}}$ include points in $\mathcal{A} \cap \mathcal{B}$. We alternate blocks of RRR and RAAR iterations throughout, with 25 iterations per block and four blocks in total, using $\beta = 0.8$. 	
\bibliography{reference}
	
\end{document}